\documentclass[11pt,preprint]{aastex}

\newcounter{magicrownumbers}
\newcommand\rownumber{\stepcounter{magicrownumbers}\arabic{magicrownumbers}}

\newcommand{\underscore}{$\_$}

\usepackage{graphicx}
\usepackage{longtable}
\usepackage{booktabs}
\usepackage{threeparttablex}

\newif\ifcolor
\colortrue

\begin{document}

\title{Structure in the 3D Galaxy Distribution:\\
II. Voids and Watersheds of Local Maxima and Minima}

\author{M.J. Way\altaffilmark{1,2}, P.R. Gazis and Jeffrey D. Scargle}
\affil{NASA Ames Research Center, Space Science Division,
Moffett Field, CA 94035, USA}
\email{Michael.J.Way@nasa.gov, PGazis@sbcglobal.net, Jeffrey.D.Scargle@nasa.gov}

\altaffiltext{1}{NASA Goddard Institute for Space Studies,
2880 Broadway, New York, NY, 10025, USA}
\altaffiltext{2}{Department of Astronomy and Space Physics,
Uppsala, Sweden}
\shorttitle{Structure in Galaxy Distribution.
II. Voids and Watersheds}
\begin{abstract}\label{Abstract}
The major uncertainties in studies of the multi-scale structure of the Universe
arise not from observational errors but from the variety of legitimate
definitions and detection methods for individual structures. To facilitate the
study of these methodological dependencies we have carried out 12 different
analyses defining structures in various ways. This has been done in a purely
geometrical way by utilizing the HOP algorithm as a unique parameter-free
method of assigning groups of galaxies to local density maxima or minima.
From three density estimation techniques (smoothing kernels, Bayesian Blocks and
self organizing maps) applied to three data sets (the Sloan Digital Sky Survey
Data Release 7, the Millennium Simulation and randomly distributed points)
we tabulate information that can be used to construct catalogs of 
structures connected to local density maxima and minima.
The resulting sizes follow continuous multi-scale
distributions with no indication of the presence of a discrete hierarchy.
We also introduce a novel void finder that utilizes a method to assemble Delaunay
tetrahedra into connected structures and characterizes regions very nearly
empty of galaxies in the source catalog.
\end{abstract}
\tableofcontents

\section{Introduction}\label{introduction}

In the last two decades an assortment of disparate density estimation techniques 
has been applied to a wide variety of data sets to characterize the distribution
of galaxies in the local universe.  From the very beginning the purely
geometrical studies were supplemented by studies of cluster luminosity functions 
\citep[e.g.][and references therein to earlier work by Zwicky]{holmberg}.
While these studies have been productive, systematic inter-comparison of the 
results continues to be problematic.  The purpose here is to address this problem
by presenting data for the construction of catalogs drawn from three data sources:
the Sloan Digital Sky Survey \citep[][hereafter SDSS]{York2000},
the  Millennium Simulation \citep[][hereafter MS]{Springel2005}
and a set of randomly distributed points. Each of these data sets were
analyzed in three different ways.  These are the same data and analysis
techniques described in the first paper in this
series \citep[][hereafter Paper I]{WGS2011}.

The first of these data sets allows elucidation of the structure of the actual
galaxy distribution.  Technology implementing fully digital, charge-coupled
device (CCD) photometric and spectroscopic observations of large areas of the sky 
has yielded a cornucopia of surveys of the local universe in the past 15 years:
e.g.  LCRS\footnote{Las Campanas Redshift Survey \citep{Shectman1996}, although
they did not actually use CCDs for their spectroscopy.},
2MASS\footnote{Two Micron All Sky Survey \citep{Skrutskie06}},
2dFGRS\footnote{The Two Degree Field Redshift Survey \citep{Colless2001}},
and SDSS.  A variety of density estimation techniques have been proposed and used
to elicit structural information from these data compendia.  Many of the methods
and the catalogs they have yielded were recently discussed in Paper I.  However,
observational surveys continue to grow larger and more elaborate, with cumulative
releases coming every 6 months to 1 year.  For example the SDSS is 
at Data Release 10 as of August 2013 \citep[][DR10]{SDSS10}. The DR10 is part of
the SDSS 3 \footnote{http://www.sdss3.org} survey scheduled to collect
data through 2014.  The complexities and shear numbers 
of both surveys and analysis methods make evaluation and interpretation
of results, and the corresponding inter-comparisons, ever more difficult.  
Even the restricted arena of density representations is replete with
different estimation techniques (of which we discuss three)
and approaches to subsequent characterization 
of the density field (\emph{cf.} \S\ref{structure_identification}).

To address this circumstance we have performed spatial structure analysis of
three directly comparable point data sets (measured, simulated, and random
galaxy positions) using three density estimation techniques (adaptive kernels,
self-organized maps, and Bayesian blocks).  We hope these analyses 
will be of use to researchers in making comparisons among 
their own methods and those described here, on a variety of redshift surveys.  
All elements of the corresponding nine-fold matrix
(3 data sets $\times$ 3 analysis methods) were described 
in Paper I.  Detailed characteristics of the data are described in Appendix A
(Section \ref{appendix_a}).

This paper describes our procedure for converting density estimates into
localized features in the spatial distribution of galaxies.
As described in Paper I this result is achieved by assembling building blocks
(tessellation cells or blocks of them) into larger structures.  A key point is
that both the localized details and global features 
that result are dependent on the principles
under which this this assembly is carried out.
Hence one of the key goals is to understand
the nature of this dependence in order to
elucidate the astrophysical meaning of 
conclusions about the origin, evolution and
current nature of the cosmic web.
In the literature these structures are typically assigned to four
classes: clusters, sheets, filaments and voids.
Our analysis leads to the point of view
that the \emph{cosmic web} \citep[e.g.][]{weygaert2009} 
is composed of a random array of structures of widely distributed shapes
not necessarily assignable to these four classes
in a straightforward way.

\section{Previous Work}\label{previous_work}

A number of recent publications have described methods for 
identifying and characterizing structure in redshift surveys and simulations.
For some developments since the summary in Paper I see
e.g. \citet{aragon_calvo_2,cautun,sousbie_2,sousbie_1,sousbie_0,falck2012,knebe,tempel}
and with respect to tessellation methods \citet{schaap,pandey,angulo}.
For a comparative study of density estimation schemes see \citet{platen2011},
and for an example of machine learning approaches see \citet{abrusco}.

Because voids were not discussed in Paper I, a brief review of the literature on
this topic is in order.  The concept of under-densities in the distribution
of galaxies and the related term \emph{void} has been around at least
since the late 1970s.  Not unexpectedly this early work was characterized
by vague definitions and uncertainties due to small sample sizes.
Some of this confusion continues to today.

\citet{chincarini_rood76} conducted one of the first observational studies indicating
the presence of voids in distribution of galaxies (for m$\lesssim$15) in the
region of the Coma Supercluster. They described the effect as a ``segregation in
redshifts," but it is now known that their survey was deep enough to see actual
voids.

The first explicit mention of voids or holes in the galaxy distribution can
probably be shared between that of \citet{gregory-thompson78} and that
of \citet{joeveer-einasto-tago77,joeveer-einasto-tago78}.
The former was published in 1978, while the latter were a pre-print from 1977
and its accepted version in 1978. The \citet{joeveer-einasto-tago77}
pre-print was also distributed in the Fall of 1977 amongst participants at
IAU Symposium No. 79 in Talinn, Estonia.  For more detail on this time
period see \citet[][p.138]{EinastoBook2014} and \citet{thompson-gregory2011}.

By the time of the 1977 IAU Symposium in Talinn, Estonia \citep{Longair-Einasto78}
voids or holes were common parlance amongst the community. Here we present a
number of examples of the relevant references.  \cite{tully-fisher78} Table II
document a \emph{void} of $>$1000 Mpc$^{3}$.
\citet{joeveer-einasto78} use the words \emph{void} and \emph{holes} in their
1978 IAU paper, and estimate on page 247 that ``Cell interiors are almost void
of galaxies; they form big holes in the Universe with diameters of 100--150 Mpc."
\citet{tifft-gregory1978} say on page 267 that ``There are regions more than
20Mpc in radius which are totally devoid of galaxies." and ``The foreground is
again very clumpy with one major void of radius close to 40Mpc."
\citet{zeldovich78} recognizes the large empty spaces (holes) discussed by others
at the conference while \citet{longair78} in his conference summary also mentions
on page 455 ``... holes which are about 10 Mpc in size and void of bright galaxies."
See also \citet{schwarzschild82} and a more recent overview of the void
phenomenon by \citet{peebles_void}.

In a pioneering mathematical study\footnote{This work was apparently
inspired by White's perception of `holes' in the galaxy distribution depicted in
\cite{gregory-thompson78}.} of probabilities that a randomly placed region of given volume 
will contain a given number (including zero) of galaxies \cite{white79} noted
that the distributions of dense structures and voids are related to each other.

According to \citet[][p.368]{martinez_saar} the first (systematic) study of voids
by \cite{einasto89} was an attempt to establish the fractal character of the
galaxy distribution.  These authors developed the \emph{empty sphere method},
thus pioneering methods to search directly for empty or near-empty volumes.

This approach threads the series of papers by \citet{el_ad_1,el_ad_2,el_ad_2a}
discussing the observational discovery of voids with the \emph{Void Finder}
algorithm (see also \citet{hoyle_vogeley} for an extension of this approach).
For automatic detection of voids in redshift surveys such as the IRAS catalog
see \citet{el_ad_3}.  The study of \citet{el_ad_2b} is of particular interest
because of its comparison of voids discovered in two independent surveys at
different wavelengths.  All of this work apparently influenced later 
methodological work close in spirit to that developed here, involving variations
on explicit search for volumes actually devoid of galaxies 
\citep{kauffmann_fairall,aikio,elyiv,tavasoli}, via nearest neighbor techniques 
\citep[e.g.][]{rojas} or ``friends-of-friends'' algorithms \citep[e.g.][]{FoF2012}.

In much other work the definition of voids is tied to local minima in the density
distribution, e.g. the Watershed Void Finder (WVF) \citep[see][and references
therein]{platen}, VOBOZ (VOronoi BOund Zones) \citep{neyrinck2005}, and 
ZOBOV (Zones Bordering on Voidness) \citep{neyrinck2008}.
These works and their concept of \emph{watersheds}
are closely related to the core idea of the HOP \citep{hop_1} algorithm adopted here.
The two classes of void finders -- based on empty volumes or local density
minima -- have their advantages and disadvantages.  The former is naturally
aligned with the discrete tessellations without smoothing
that characterizes our previous work in Paper I.
See also the recent works by \cite{neyrinck2013} and \cite{nadathur2014}.

\cite{schmidt} deal with voids in simulations, comparing methods based on finding
empty regions of space (within observational limits and selection effects in the
survey) against those based on density estimation followed by identification of
density minima.  They also include two different void finder algorithms with and
without predefined constraints on shape.  Extensive studies of the structure and
dynamics of voids \citep{aragon_calvo_3,aragon_calvo_4} argue for the existence
of a hierarchical distribution of voids in the context of the \emph{cosmic web} 
and \emph{cosmic spine} concepts \citep[see also][]{aragon_calvo_1,aragon_calvo_2}.
While the term \emph{hierarchy} is commonly invoked for both voids and structures
no evidence for discrete levels is found in analysis of survey data.  We feel
that the distributions of sizes of structures are better described as continuous
and multi-scale, not hierarchical (cf. the comments in Section \ref{conclusion}.)

A number of recent works have investigated the statistics and stacking of voids
and their importance for various environmental and other cosmological issues
\citep{hahn_1,hahn_2,aloisio,gaite,paranjape,lavaux2010,einasto_b,pan,
einasto_a,bos,lavaux2012,bolejko,zaninetti,varela,sutter,jennings,
beygu,krause,ceccarelli,hamaus,ricciardelli,hamaus2014}.
A comparison of void catalogs and detection methods applied to the MS data is
found in \citet{colberg}; see also \cite{knebe2011}.
And more generally, powerful methods of point process theory \citep{daley,lowen},
stochastic geometry \citep{snyder}, discrete Morse theory \citep{sousbie_1,sousbie_0},
computational \citep{deberg,preparata} and combinatorial
geometry \citep{edelsbrunner_1}, and wavelet-like transforms \citep{leistedt2013}
are currently being used to explicate multi-scale structures in the galaxy
distribution \citep{weyg_a,weyg_b,weyg_c,sousbie_2,sousbie_1,sousbie_0,park,hidding}.
Especially interesting are the prospects for studying voids via gravitational
lensing effects \citep{amendola1999,higuchi,melchior2013}.

\section{Identification of Structures}\label{structure_identification}

\begin{figure}[htb]
\includegraphics[scale=.60,angle=90]{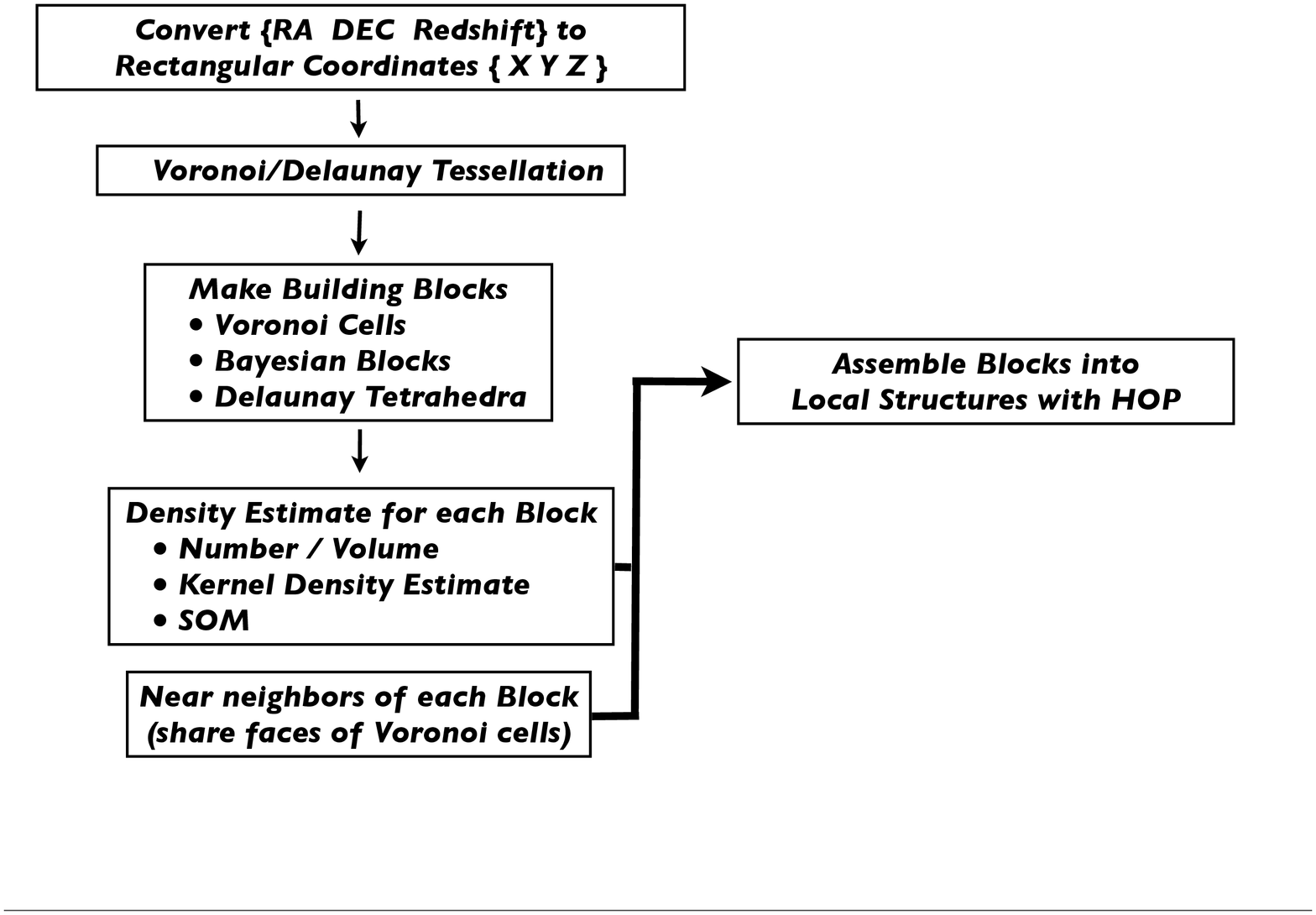}
\caption{Flow chart for the analysis procedure.  Processes on the left were
described in Paper I; assembling building blocks into structures is the main
topic of the current paper.  The tessellation techniques and self-organizing
maps (SOM) are as in Paper I.  HOP \citep{hop_1} is the assembly algorithm
adopted here.}\label{figure01}
\end{figure}

The grand challenge is to produce scientifically useful characterization of a
density field derived from a galaxy survey or a computational dark matter
simulation.  One approach is to study statistical quantities averaged over the 
hole data sample.  Examples include estimation of correlation functions 
\citep{mcbride2011,valageas2012,mueller2012}, power spectra
\citep{tegmark2006,jasche2010,neyrinck2009}, and global topological information 
\citep{shandarin2004,gott,james2009,weygaert2010,sousbie_1,sousbie_2,einasto_c}.
Here instead we develop an alternative approach, namely identification of
specific local features of the density distribution, as outlined in
Figure \ref{figure01}.

In Paper I the steps on the left side of the figure yielded density surrogates
for individual galaxies or small sets of them.  The current paper addresses the
assembly of these building blocks into structures.

It is under-appreciated that the results of any such analysis are very dependent
on the methodology used.  Especially strong is the dependence on the assembly
procedure, but all of the choices represented in Figure \ref{figure01} have their
effects.  There is a plethora of definitions of structural classes and approaches
to detecting them, using a variety of positional, photometric and morphological
information.  In particular some detection algorithms, such as those invoking
prior information about galaxy colors or cluster symmetry, naturally favor
detection of structures more nearly satisfying these assumed properties.
This rather murky situation raises questions.  Are there intrinsic distinct
well-defined structural classes?  If there are, can we uncover their nature in
objective ways not unduly influenced by methodology and prior assumptions?
Since structures can have, macroscopically speaking, four possible dimensions
(0, 1, 2, and 3) the corresponding shape classes -- clusters, filaments, sheets,
and voids -- seem natural.  This classification scheme has been adopted by the
community, with some recognition that shapes are somewhat randomly distributed
in and between these categories.  In any case it is important to exercise 
care in the interpretation of structural results.  Comparison of observed and
simulated data using identical analyses is inherently less ambiguous, 
but even such comparative studies depend on the nature of the analysis.

Our work seeks to avoid some of this murkiness by using purely geometrical
information derived from galaxy positions, and in ways not tuned to emphasize
any particular shape.  The approach here is to characterize structure over a
range of scales, i.e.  \emph{multi-scale structure} (a more precise term than
the commonly used \emph{large-scale structure}).
We aim to make maximal use of the information in the data, but with neither
prior shape constraints nor account of geometrically extraneous systematics 
such as the \emph{red sequence} in clusters \citep{gladders,redmap_1,redmap_2,redmap_3}.
We adopt what is arguably the simplest possible definition: \emph{a structure is
the watershed of a single critical point} -- that is of a local density maximum
or minimum (see Section \ref{hop} for specifics).  This choice rules out
structures with two peaks e.g., but if desired these could be sought during a
post-processing with some kind of merger criterion.  Identification of voids via
local density minima is supplemented with a novel void finding procedure in
Section \ref{delaunay_void_tracer}.  In a further bid toward objectivity and
parameter freedom we use tessellation techniques so that the scale on which
these maxima are determined is automatic, data-adaptive, and not predefined.
The full definition of structures then requires a prescription for what to attach
to the local maxima or minima.  Here we assemble structures out of elementary
building blocks as outlined in Section 4.3.3 of Paper I, avoiding arbitrary
choices through the use of the parameter-free HOP algorithm of \citet{hop_1}.

\subsection{The HOP Algorithm}\label{hop}

The rest of this paper is devoted to the process of assembling structures out of
building blocks (cf. the right-hand side of Figure \ref{figure01}).
The next subsection describes a general algorithm for assembling local structures
in any data representation consisting of these three elements:

\begin{itemize}
\setlength{\itemindent}{.5in}
\item[(1)] A set of discrete entities, called \emph{objects}
\item[(2)] The value of a function $\bf{f}$ for each object
\item[(3)] Adjacency information among the objects
\end{itemize}

\noindent We refer to $\bf{f}$ as the \emph{HOP function}.
For the computations only the last two items matter, as the algorithm makes no
reference to the identity of the objects.
In continuous Morse theory \citep{milnor} the \emph{Morse function} -- the analog
of our $\bf{f}$ -- must be infinitely differentiable;
in the discrete theory of \cite{forman} the corresponding function must satisfy
some similarly delicate conditions.  Here in contrast the HOP function is
essentially arbitrary since it only needs to provide an ordering of the objects.
Hence the only condition on ${\bf f}$ is that no two objects are assigned the
same value; violations of this constraint can be fixed in a trivial way.

In some of the cases reported here the objects are individual galaxies,
with the HOP function given by a density value assigned to each one (e.g
via Voronoi cell volumes, kernel density estimates,
or SOM class identifiers related to density).
In another -- the case of Bayesian Blocks \citep{scargle_vi} -- the objects are
connected sets of adjacent galaxies with the appropriate density estimate (number
of galaxies in the block divided by its volume).  In the final example considered
here, where the objects are the Delaunay tetrahedra in the tessellation of galaxy
positions, the definition of ${\bf f}$ requires some thought, as elaborated in
Section \ref{delaunay_void_tracer}.

For the adjacencies referred to in item (3) above we use those defined by the
Voronoi tessellation itself, as follows:
two objects, either individual Voronoi cells or blocks thereof, 
are deemed adjacent if they intersect at a common 
face.\footnote{Objects being joined by an edge in the Delaunay tessellation is
equivalent to this condition (but not to other possible definitions such as
sharing Voronoi edges or vertices in lieu of faces).  These adjacencies are
easily established from information supplied by most data analysis systems, such
as the n-dimensional MatLab routine \emph{voronoin}, namely identities of the
vertices of each Voronoi cell.  In finding adjacencies it is very useful to first
compile a list of all galaxies whose Voronoi cells touch each vertex.} 
This construct can be viewed as defining nearest neighbors in a data-adaptive
fashion, with no a priori restriction on the number of neighbors. 
It thus conveys local information regarding the distribution of galaxies 
more efficiently than say nearest-neighbors with a pre-defined number of neighbors.
Similarly two Delaunay tetrahedra are considered adjacent if they share a common
triangular face.

Identification of watershed structures in a spatial distribution of objects 
is conveniently implemented using the group-finding algorithm HOP \citep{hop_1}.
Here by the term HOP we mean only the group-finding step (their Section 2.1) 
distinguished from data smoothing and merging of groups discussed elsewhere in
their paper.  \cite{aubert} describe a variant of this post-processing,
called {\bf AdaptaHOP}; their approach differs in many ways from ours, such as
regarding the sampling as noise to be smoothed over, and has procedures that
generate hierarchical leaves in a tree.  See also \cite{Springel1999} for a
discussion of a related algorithm called {\bf SUBFIND}.  Motivated
by the requirements for analysis of massive cosmological data sets, \cite{hop_2}
deals with computational parallelization issues.  \cite{turk} provide a toolkit
that contains an implementation of HOP.

This algorithm is quite general. For any given HOP function and adjacencies 
defined for each object in set $\bf{S}$, it yields a partition of $\bf{S}$
into groups with these properties:

\begin{enumerate}
\item[(a)] The elements of the partition, here called \emph{groups}, are sets of objects from $\bf{S}$.
\item[(b)] There is one such connected group for each local maximum of $\bf{f}$.
\item[(c)] In a given group $\bf{f}$ decreases monotonically away from the maximum.
\item[(d)] Every object is in one and only one of the groups (i.e., the groups partition the space).
\item[(e)] The partition is unique and  parameter-free.
\end{enumerate}

In short HOP identifies all of the local maxima of $\bf{f}$ and the connected
structures flowing from them; together these are the discrete analog of the
so-called \emph{descending manifolds}  --  mountain peaks plus their watersheds.
It can identify structures of any shape -- containing arbitrary mixtures of
convexities and concavities, possibly even failing to be simply connected.
The underlying idea of HOP is a simple hill climbing prescription.
It iteratively associates each object with neighbors that have larger values
of $\bf{f}$ according to this formulation:

\begin{center}
\textbf{The HOP Algorithm \citep{hop_1}}
\end{center}
\begin{itemize}
\item[\underline{Given:}]A set $\bf{S}$ of {\bf N}  spatially distributed data objects $\bf{o_i, i = 1, 2, \dots N}$
\item[(1)]Establish an index array $\bf{I = \{ 1, 2, 3, \dots N \} }$ for any convenient ordering of the objects.  
\item[(2)]Assign a value of $\bf{f}$ to each object.
\item[(3)]For each object identify all objects adjacent to it, i.e. its spatial
neighbors as defined earlier in this section.
\item[(4)]For each set consisting of an object and all of its neighbors, find the object 
with the largest value of $\bf{f}$.
\item[(5)]Iterate as follows:
\begin{itemize}
\item[(a)] For $\bf{i = 1, 2, \dots N}$: 
\begin{itemize}
\item[(i)] Let $\bf{j_i}$ be the  current index value in position $\bf{i}$ of $\bf{I}$ 
\item[(ii)] In I replace $\bf{j_i}$ with with that found in (4) for object
$\bf{j_i}$ (not that for object $\bf{i}$)
\end{itemize}
\item[(b)] Repeat (a) until no index value changes
\end{itemize}
\item[(6)] Set ${\bf K} =  {\bf I}$ with duplicate values removed. 
\label{hop_algorithm}
\end{itemize}

\noindent Eliminating the duplicate values in the converged $\bf{I}$   
yields a set of indices $\bf{K}$ pointing one-to-one to each of the local density
maxima -- objects denser than all of their neighbors.  Each of the objects ends
up pointing via the converged ${\bf I}$ to one and only one 
of these maxima (i.e. to one of the values in ${\bf K}$).
This property generates for each local maximum a connected structure,
consisting of a set of paths connecting adjacent galaxies along which
the density is monotonic. These structures are much like the 
\emph{watersheds} defined in image processing and
many of the cosmic web algorithms referenced above.  

In short this algorithm uses a simple hill climbing procedure to find a unique
partition of the objects into groups, one associated with each of the local
maxima.  Alternatively, by jumping instead to the neighbor with the smallest
value of $\bf{f}$ in Step 4 HOP can find basins of attraction for all of the
local \emph{minima} instead.  We usually refer to structures associated with
local maxima as \emph{groups}, rather than clusters, since they may e.g. be
filamentary or sheet-like.  With a similar freedom with standard terminology, 
drainage basins associated with local minima can be called \emph{voids},
although we distinguish these structures from the empty collections of Delaunay
tetrahedra discussed in Section \ref{delaunay_void_tracer}.

The following MatLab code fragment uses vector operations to implement the
iteration, given two arrays initialized as follows:

\verb+index = [1, 2, ... N]+ contains the initial indices
of the objects (taken in arbitrary order)

\verb+id_max_neighbor+ contains indices of the neighbors found in step 4

\noindent
\begin{verbatim}
 while 1
       index_new = id_max_neighbor( index );    % Each object hops to highest neighbor
       id_change = find( index_new ~= index );  % Locate index changes
       if isempty( id_change );break;end        % If no index changes escape while loop
       index = index_new;                       % Implement changes due to hops
 end
 \end{verbatim}
\noindent

We take the objects in $\bf{S}$ to be individual galaxies, blocks containing
several galaxies, or Delaunay tetrahedra -- attached to which are values
of $\bf{f}$.  This function is given by the corresponding KDE or BB density
estimates, classes from the SOM method, or derived from the sizes of the Voronoi
cells or Delaunay tetrahedra.  

The following points elaborate some details of the algorithm and our application
of it.

\begin{enumerate}

\item Choices for the definition of the neighbor relation in step 3, and indexed
by \verb+id_max_neighbor+, include densest neighbor (to find manifolds descending
from local maxima) and least dense neighbor (to find manifolds ascending from
local minima).

\item Throughout this discussion there is no explicit mention of the
dimensionality of the data.  One of the beauties of the HOP algorithm is that it
applies to spaces of any dimension.  Here contact with the dimension of the data
arises only in the definition of adjacency, which we compute from the Voronoi
tessellation of the 3D galaxy positions.  But once the adjacencies are assigned
dimension is completely irrelevant.

\item The unique output of this algorithm is independent of the order of the
initial indexing (1) or the order in which the objects are considered in
step (5), modulo an inconsequential re-ordering of the output groups.

\item The iteration can be carried out in other ways than shown explicitly above,
e.g. by following individual objects to their final destinations, rather than the
parallel procedure in (5)(a).  Such path tracking is of use for constructing
analogs of topological saddle points, not discussed here.

\item If two or more objects are assigned identical values, rare except in the
case of the discrete SOM class identifiers, there may be a dependence on the way
the resulting ambiguity is resolved. 

\item After the preprocessing represented by the initial steps (1) - (4),
iteration (5) is guaranteed to converge rapidly because of the monotonic nature
of the bounded upward jumps, which are the source of the name HOP.

\item All that matters is the ordering of the function values,
so $\bf{f}$ can be replaced with ordinal numbers. i.e.
integers indexing the array $\bf{f}$ in increasing order.\label{item_7}

\item The first step of the ZOBOV algorithm \citep{neyrinck2008} and most Morse
theory algorithms is based on what is essentially the same as HOP.

\item There is no loss of information due to smoothing
in the process of assembly of objects into structures,
although Bayesian Blocks can be thought of as a form of smoothing
(more properly ``chunking'') in preprocessing.

\item HOP is a major simplification, sidestepping much of the complexity of Morse
theory (continuous or discrete) and persistency concepts that characterize 
modern topological data analysis (cf. references in Section \ref{previous_work}).

\begin{itemize}

\item Nonetheless the results presented here compare favorably to those from more
elaborate algorithms, for example based on discrete Morse theory.  The essential
difference is that small structures, discarded by others because they are not
\emph{persistent} as some parameter is varied, we regard as conveying important
information and are therefore retained in our analysis

\item One concern is that our resulting structures might extend 
from their defining local maxima down to low density levels
that might be better assigned to local minima.
Some aspects of topological data analysis address this issue by
truncating structures, utilizing saddle points and intersections of
ascending and descending manifolds.  However, the structures
found here without such procedures do not seem to have
any pathological features, such as tentacles extending far from
the defining critical point.

\end{itemize}

\end{enumerate}

\noindent
Before showing examples of structures determined with HOP, the next section
describes some considerations relevant to another way to detect voids.

\subsection{Delaunay Tetrahedra as Void Tracers}\label{delaunay_void_tracer}

As described in Section \ref{introduction} the so-called \emph{voids} 
in the galaxy distribution have been the subject of considerable study.
These features are informative regarding the multi-scale structure of the
Universe, just as are dense structures.  A variety of definitions of voids,
and detection methods keyed to the defining characteristics, provide different
views of both individual and overall structures.  As discussed above in
Section \ref{previous_work} some detection methods focus on local density
minima; others locate volumes of space empty of galaxies within the 
limits of the survey or simulation, with no explicit reference to 
local minima in a continuous density representation.  We here develop an approach 
of this latter kind that we believe is novel in its explicit use of
Delaunay cells as the building blocks for the voids.

\begin{figure}[htb]
\includegraphics[scale=.9]{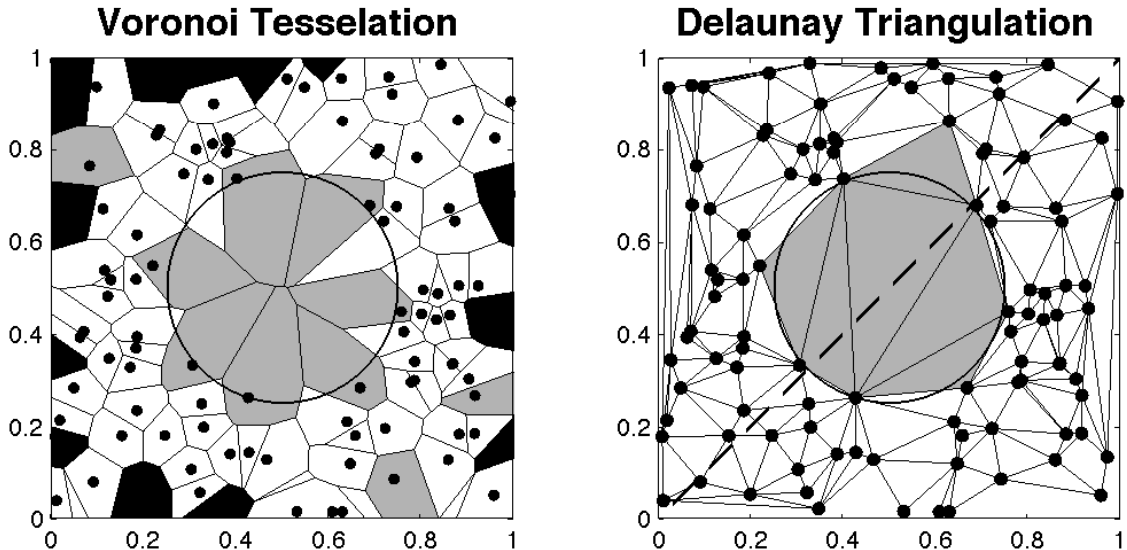}
\caption{Tessellations for a synthetic 2D circular void.  Points shown as dots
are randomly and uniformly distributed in the unit square excluding a circle of
radius ${1 \over 4}$. Left: Voronoi cells with areas $> .023$ and $> .05$ are
shaded gray and  black.  Right: Delaunay triangles with areas $> .019$ shaded gray.
(The thresholds were chosen to roughly optimize the void representations.)}
\label{figure02}
\end{figure}

Delaunay and Voronoi tessellations are \emph{duals}\footnote
{This concept refers to several relations. See \citep{spatial_tessellations}
for details.} to one another, each partitioning the data space into
small sub-volumes in different ways but elucidating similar spatial information.
We have seen that cells in a Voronoi tessellation of galaxy positions are good
building blocks for constructing a representation of the corresponding density
field.  However Delaunay tessellation is much more effective than Voronoi
tessellation for void detection.  The toy example in Figure \ref{figure02}
compares their responses to an artificial empty region in a set of otherwise
random 2D points.  The strategy is to find an objective way to identify a set of
cells approximating the void, for example by selecting those cells larger than
some adopted size threshold.

Several problems beset Voronoi tessellation's partial success  in the left panel.
Identification of void cells is relatively complicated.  Both upper and
lower thresholds are required, the putative void cells shown in gray
lying in between.  The cells above the upper threshold
(black) are oversized due to edge effects  (cf. Paper I)
and those below the lower threshold (white) are in the denser non-void region.
Typically there is no range of cell areas that includes all the cells in the void 
and only those cells -- that is rejecting both edge and extra-void cells. In the
left panel of Figure \ref{figure02} three obvious edge cells are incorrectly
denoted as void cells (grey).  Adjusting the upper area threshold to correct
this mistake eliminates some true void cells.  Even the best Voronoi coverage
does not provide a very exact representation of the void's shape and extent.
Finally, since they must contain a data point, the identified void cells 
not only extend outside the true circular void but they carry a nontrivial
density, namely unity divided by the cell area.  On the other hand in the
Delaunay triangulation (Figure \ref{figure02}, right panel) the circular void is
well represented using a single area threshold yielding a small number
of triangles, each of which is empty and can be interpreted as carrying zero
density.  

\begin{figure}[htb]
\includegraphics[scale=.99]{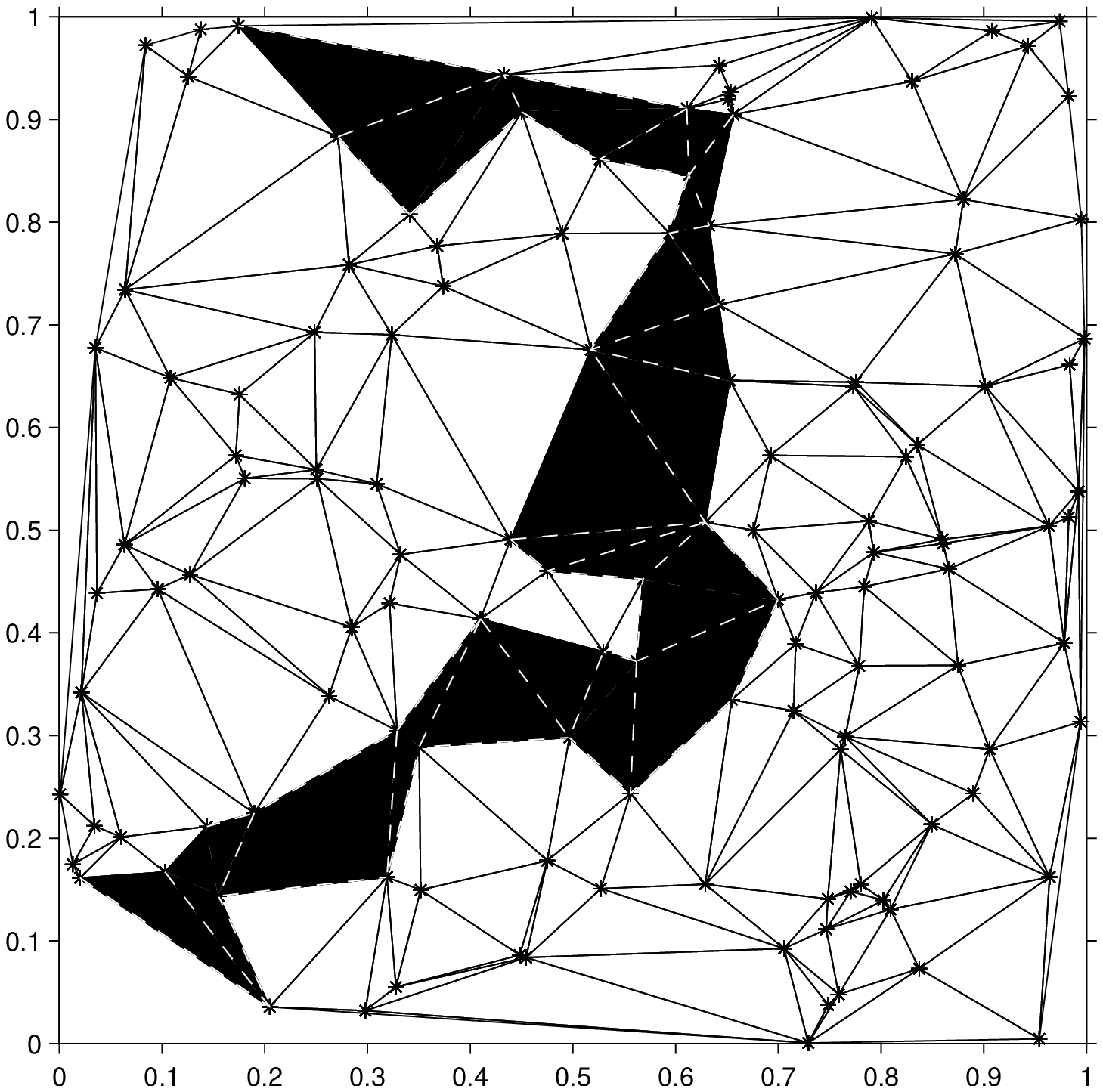}
\caption{Delaunay triangulation of 132 random point in the unit square.  The set
of shaded triangles comprises a connected region empty of points but with a shape
dictated by the rather arbitrary choice of which triangles to paste together.}
\label{figure03}
\end{figure}

Furthermore sets of several contiguous Delaunay tetrahedra typically make up
structures devoid of galaxies. Figures \ref{figure02} and \ref{figure03} show
quirky 2D examples of this fact. In the right panel of Figure \ref{figure02} the 
dashed line running diagonally between the lower-left and upper-right corners
does not obviously define a structure of any interest, but in fact the set of 
Delaunay triangles which it intersects is a void in the form of jagged polyhedron
not containing any of the points.  Figure \ref{figure03}  demonstrates the same
thing for triangles intersected by a continuous curve (not shown) in the plane.
Neither of these shapes are what one thinks of as reasonable voids, but as
discussed in Section \ref{noise} this does not mean that they are somehow not
real.  In principle collections of adjacent tetrahedra are not necessarily empty
of galaxies (cf. Section \ref{empty}) but in the analyses presented here
they always are.

The 2D toy examples in Figures \ref{figure02} and \ref{figure03} were chosen for
ease of visualization but these conclusions are even more definitive in 3D: many
paths similarly define snake-like connected configurations of empty Delaunay
tetrahedra (cf. Figure \ref{figure03}).  Detected void structures are very
sensitive to the assembly process, as was evident in the Aspen-Amsterdam void
finder comparison project. The void volume fractions with the various void
finders reported in column 4 of Table 2 of \citet{colberg} range from 0.13 to 1.0.
The last number, due to Platen and Van de Weygaert, means the entire data space is 
represented as a single highly convoluted but empty polyhedron. With a goal of
identifying coherent low density structures an obvious scheme is to start from
locally largest Delaunay tetrahedra and apply criteria for attaching one or more
of the 4 face-sharing neighbors, perhaps based on something like size, shape or
distance.  To avoid subjectivity of such ad hoc procedures we use HOP's
prescription for partitioning the set of tetrahedra comprising the Delaunay
tessellation (Section \ref{hop_structures}).  To do this we need a new definition
of the function $\bf{f}$ in algorithm \ref{hop_algorithm}, since a surrogate for
galaxy density is not appropriate for void finding. In order to represent the
degree of emptiness we take $\bf{f}$ equal to tetrahedron volume.  In much the
same way that small Voronoi cells correspond to large density, large Delaunay
cells correspond to a large degree of emptiness.\footnote{Tessellation
ameliorates dependencies on the size of the region sampled.  `` ... if you want
to measure the density of biomass in a treetop, you have to choose a window of
maybe a cubic foot.  Ten times less and you sample either a single leaf or a
blob of air.  Ten times more and you have almost reduced the tree to an
operational point.'' \citep{koenderink}.  Tessellation turns this dilemma on its
head: the data points adaptively fix both the size and location of the windows.}

\subsection{Ambiguity, Uncertainty, Noise and Persistence}\label{noise}

Quantifying uncertainty in data analysis requires careful assessment of any process
affecting the signal of interest, either randomly or systematically, 
anywhere along the entire chain leading from raw measurements to the final estimate. 
In addition subsequent interpretation must allow for dependencies of the results
on the analysis method.  Accordingly at least the following issues need to be considered 
in assessing the cosmological significance of the present analysis:
\begin{enumerate}
\item[(1)] Errors in measured sky positions and redshifts of individual galaxies.
\item[(2)] The effect of random motions on estimated distances.
\item[(3)] Sampling bias connected with fiber collisions.
\item[(4)] Distortion of tessellation cells near edges of the data space.
\item[(5)] The initial random field of density perturbations.
\item[(6)] The finite number of galaxies forming randomly in the evolving density field. 
\item[(7)] Sampling a small subset of the galaxies that have formed in the given volume.
\item[(8)] The variety of distinct but equally justifiable definitions of structures.
\item[(9)] The variety of different analysis methods.
\item[(10)] The variety of different selections of input data.
\end{enumerate}

Many of these items are either noise (to be removed, diminished, or otherwise
accounted for) or signal, depending on the context.  The following discussion
addresses this distinction, for the listed items, as dictated by our goals.
The direct observational errors in (1) are described in \citep{Blanton05,SDSS7},
and include both small approximately normally distributed errors and larger
outliers.  In a nutshell the sky-positional errors are quite small on the scale
of interest here, and the random distance uncertainties derived from redshift 
errors are on average much less than $\approx$ 0.5 Mpc.  Here we do not try to
remove redshift distortions (2), reserving their treatment to post-factor
examination of the shapes of dense clusters. The sampling bias issues connected
with fiber collisions (3)  and Voronoi cells near edges (4) were discussed
in Paper I. For the present purposes all of these errors can be assumed to 
be small in magnitude and, due to the inclusion of many individual galaxies,
should not  substantially impact the overall results.

The remaining entries in the above list require more discussion. Some are part of
the astrophysical signal of interest.  A large amount of research has been
devoted to issues of uncertainty in computational topology \citep{edelsbrunner_3}
and topological data analysis \citep{zomorodian}.  This work has focused on 
simplifications realized by discarding or consolidating less important features
yielded by various analysis schemes.  Examples of goals of this approach are 
amelioration of noise, especially  \emph{discreteness noise}; reduction of
complexity and memory requirements; and promotion of better visualization and
understanding of structure revealed by removing extraneous details. This
methodology requires quantification of importance, ideas for which range from
simple size criteria to the rather complex notions of \emph{topological persistence}
\citep{edelsbrunner_1,edelsbrunner_2,gyulassy,edelsbrunner_3,gerber,carlsson,chen}.
Persistence methods postulate that the importance  of a feature is measured by how
long it is present as some parameter is scanned over a range of values.  A
quantitative link from persistence to probability may be obtained using bootstrap
methods \citep{marzban,chazal,fasy}.  More recently the persistence concept has  
made its way into astronomical  applications \citep{sousbie_1,sousbie_2,sousbie_0,cisewski}
resulting in use of the term \emph{the persistent cosmic web}.

However, with our goal of characterizing the complete range of multi-scale
structure these methods discard some of the very information we seek.  Modern
cosmology posits that structure in the Universe started as spatially random
density fluctuations.  Our Universe evolved deterministically from this single
set of initial conditions; this process involves nothing like an ensemble of 
realizations of a random process (as in errors of observation).  Accordingly we
consider items (5) and (6) signal, not noise.  However in other contexts, such as
dark matter simulations, discarding small structures as unimportant consequences
of initial spatial randomness may be useful.  Item (7), sometimes called
\emph{discreteness noise}, is inherent to data consisting of a limited number of 
points draw from an unknown distribution.  Appendix B of \cite{liivamagi} gives
a detailed error analysis of this concept based on the Poisson model of 
\cite{peebles_2}.  But for reasons similar to those discussed with regard to
(5) and (6), we also regard (7) as part of the astrophysical signal of interest.
(Nevertheless the random Poisson data we have included may be of use in other
contexts where noise abatement may be useful.) Our Appendix B contains some
further remarks about potential effects of what is often called topological noise.

While any of these last three factors, (5)-(7), are possible justifications for
simplification using topological persistence or related measures, neither is
actually a source of uncertainty about the reality and nature of multi-scale
structure in the current Universe derived from a given redshift survey.
The distribution of structures derived from discrete samples provides
information about initial fluctuations and their subsequent evolution.
Therefore removing or smoothing away small scale structures is at worst
discarding useful cosmological information;  at best it makes the 
conclusions dependent on postulated models for the relevant physical processes.
For this reason, and because the goal here includes geometrical and not just
topological analysis, we do not employ any of the simplification procedures cited
above.  But in other contexts such as global analysis (e.g. estimation of a few
summary topological statistics, such as genus, Minkowski functionals, or Betti
numbers) it may be reasonable to regard scatter about a smooth correlation
function or within realizations from different initial data as noise.  In such
cases countermeasures such as topological persistence techniques may be justified.

The largest source of ambiguity in multi-scale structure is the strong dependence 
of analysis results on analysis methodology, and the fact that there is no one
correct methodology or definition of structures -- cf. items (8)-(10).
For example, we saw that by merely adjusting the halting criterion for assembling
Delaunay tetrahedra into voids (Section \ref{delaunay_void_tracer}), the output 
of voids ranges from a single void encompassing the entire space
(cf. Figure \ref{figure03}) to a void for each tetrahedron.  The question is not
where between these extremes the truth lies but what representations 
provide the most useful information -- for example in the comparison of observations
and simulations.  Better yet, it can be very fruitful to study structural
representation as a function of methodological assumptions and values of
parameters of the analysis.

The HOP results that follow are examples of convenient representations using a
simple notion of \emph{attaching to an elementary structure the neighbors that it 
dominates} -- as in the definitions of Voronoi cells, Bayesian blocks, and
groups of building blocks that thread this paper.  While a fairly natural
construct, this is by no means claimed to be better or more fundamental than 
any others.

\subsection{Structures Obtained with the HOP Algorithm}\label{hop_structures}

Now turn to some examples of the identification of spatial structures using the
HOP algorithm to assemble the elementary \emph{objects} or building blocks (i.e.
Voronoi cells, Bayesian Blocks or Delaunay Tetrahedra) into a unique set of
connected structures.  Each such structure descends or ascends monotonically
from one of the \emph{critical objects} -- local maxima or minima of the adopted
density or voidness function. These peaks -- for example each Bayesian block
denser than all its face-adjacent neighbors -- can be easily identified by
direct search but are also automatically produced by the HOP algorithm.
The structures attached to the peaks are analogous to their watersheds.
Such structures could be classified in one way or another (e.g. in the four
customary classes: clusters, filaments, sheets and voids, macroscopically of
dimensionality 0, 1,  2 and 3 respectively) but their shapes are widely
distributed in shape-space and they do not fit cleanly in discrete clusters
of shape parameters.

The sole information needed for each galaxy consists of two items, the first
being the value of \emph{the HOP function} ${\bf f}$ -- typically a density
estimate or its surrogate.  It is natural for tessellation-based studies 
to take as the density of an object the number of galaxies in it divided
by its volume.  The reasoning for individual Voronoi cells is straightforward:
small cells occur in crowded regions where the cell size is small.
This relation intuitively supports the idea that the reciprocal of a cell volume
is a reasonable surrogate for local density at or near that cell.  In addition
this construct can provide an unbiased estimate of local density \citep{platen2011}.
The fact that only the relative order of the densities matters (item \ref{item_7}
in the list of properties of HOP in Section \ref{hop}) is further
protection against bias effects.  Correspondingly the density we assign in the
case of Bayesian Blocks is the number of galaxies in the block divided by its
volume, the latter defined as the sum of the volumes of the cells making up the
block.  In the KDE case we evaluate the  estimated continuous density
field at the position of each galaxy.  One of the SOM parameters is taken
as a rough density surrogate (Paper I).

The other item necessary is a list of adjacent neighbors for each galaxy.
As indicated earlier in Section \ref{hop} for the Voronoi-based tessellations, 
we take two objects $A$ and $B$ (cells or blocks) to be adjacent to each other
if and only if there is at least one pair of Voronoi cells, one member of the
pair in $A$ and the other in $B$, which share a common face.  Here we take
advantage of the natural definition of a data-adaptive number of near neighbors
that Voronoi tessellation provides. The KDE algorithm does not determine
neighbors, and we simply impose the adjacency information copied from the
Voronoi tessellation. Two Delaunay tetrahedra are considered adjacent if and
only if they share a common triangular face.

Note that Voronoi cell volume is not a property of a single galaxy but is
determined by its propinquity to its neighbors; hence information from distances
to other galaxies is represented in both the cell volumes and the identities
of neighboring cells.

Table \ref{table01} summarizes for the SDSS data some of the basic properties of
the collections of structures resulting from five choices for the building blocks
for structures.  In higher dimensional Bayesian
blocks \citep{jackson_higher_dimension} one constructs a 1D array consisting of
ordered values of a cell variable.  In Paper I this quantity was taken to be the
volume of the Voronoi cell.  Since HOP more naturally operates 
on density we also redid the whole Bayesian Block analysis 
of Paper I, this time using density as the cell variable instead of volume.
These two analyses are listed in the first two rows of the table,
showing that there is not a large difference in the number of structures identified.  
The nature of the HOP input for the other 3 cases are described by the
corresponding entries in the first two columns of the table.  The second column
indicates the definition of ${\bf f}$, taken to be the density of galaxies within
a Bayesian Block or Voronoi cell, KDE density, or SOM class.  The objects fed to
the HOP algorithm (Column 1) are individual galaxies except in the first two
cases, where they are collections of galaxies in blocks.  The third column
indicates the number of objects input to the algorithm.  The last two columns
give the number of structures, or groups of galaxies, associated with density
maxima and minima.
\begin{table}[htdp]
\caption{Statistical Summary of SDSS Structure Collections.  $V_{block}$ and
$V_{Voronoi}$ are the volumes of the blocks and cells, respectively, and
$N_{gal}$ is the number of galaxies in a block. There are 146,112 galaxies before
the fiber collision and edge cuts.}\label{table01}

\begin{center}
\begin{tabular}{|  l    |  c | r |  l  c |  }
\hline 
Objects &   ${\bf f}$ & Num. of Objects & ~ ~ ~  Num.  of  Structures &  \\ 
              &      &   &   Maxima  &  Minima  \\
\hline \hline
Bayesian Blocks (volume) & $N_{gal} / V_{block}  $ & 41,672 & 7,517  &  752  \\ 
Bayesian Blocks (density) & $N_{gal} / V_{block}  $ & 46,491 & 11,273 &   178  \\ \hline
Galaxies & KDE  & 133,991 & 6,615 & 1,796 \\  \hline
Galaxies & $1 / V_{Voronoi}$  & 133,991 & 10,414 & 1,032  \\ \hline
Galaxies & SOM class & 133,991 & 2,076 & 10,516 \\ 
 \hline
\end{tabular}
\end{center}
\label{table02}
\end{table}

The following two tables record similar information for the
other two data sets, the Millennium Simulation, and 
independently and randomly distributed points, respectively.

\begin{table}[htdp]
\caption{Statistical Summary of MS Structure Collections.  $V_{block}$ and
$V_{Voronoi}$ are the volumes of blocks and cells, respectively, and $N_{gal}$
is the number of galaxies in the blocks. There are 171,388 galaxies before the
fiber collision and edge cuts. }\label{table03}
\begin{center}
\begin{tabular}{|  l    |  c | r |  l  c |  }
\hline 
Objects &   ${\bf f}$ & Num. of Objects & ~ ~ ~  Num.  of  Structures &  \\ 
              &      &   &   Maxima  &  Minima  \\
\hline \hline
Bayesian Blocks (Volume) & $N_{gal} / V_{block}  $ & 54,850 & 8,848  &   1,036  \\ 
Bayesian Blocks ($\rho$) & $N_{gal} / V_{block}  $ & 57,305 & 12,023 &   404  \\ \hline
Galaxies & KDE  & 148,927 & 9,859 & 9,767 \\  \hline
Galaxies & $1 / V_{Voronoi}$  & 148,927 & 12,618 & 532 \\ \hline
Galaxies & SOM class & 148,927 & 11,603  & 12,979  \\ 
 \hline
\end{tabular}
\end{center}
\label{ms_statistics}
\end{table}

\begin{table}[htdp]
\caption{Statistical Summary of Poisson Structure Collections.  $V_{block}$ and
$V_{Voronoi}$ 
and $N_{gal}$ is the number of galaxies in the blocks. There are 144,700 points
before the fiber collision and edge cuts. }\label{table04}
\begin{center}
\begin{tabular}{|  l    |  c | r |  l  c |  }
\hline 
Objects &   ${\bf f}$ & Num. of Objects & ~ ~ ~  Num.  of  Structures &  \\ 
              &      &   &   Maxima  &  Minima  \\
\hline \hline
Bayesian Blocks (Volume) & $N_{gal} / V_{block}  $ & 30,218 & 6,823  &  3,090  \\ 
Bayesian Blocks ($\rho$) & $N_{gal} / V_{block} $ & 42,471 & 11,009 &  2,395 \\ \hline
Galaxies & KDE  & 131,832 & 2,313  & 2,765  \\  \hline
Galaxies & $1 / V_{Voronoi}$  & 131,832 & 10,009 & 3,374  \\ \hline
Galaxies & SOM class & 131,832 & 8,558 & 11,522 \\ 
 \hline
\end{tabular}
\end{center}
\end{table}

\begin{figure}[htb]
\includegraphics[scale=1]{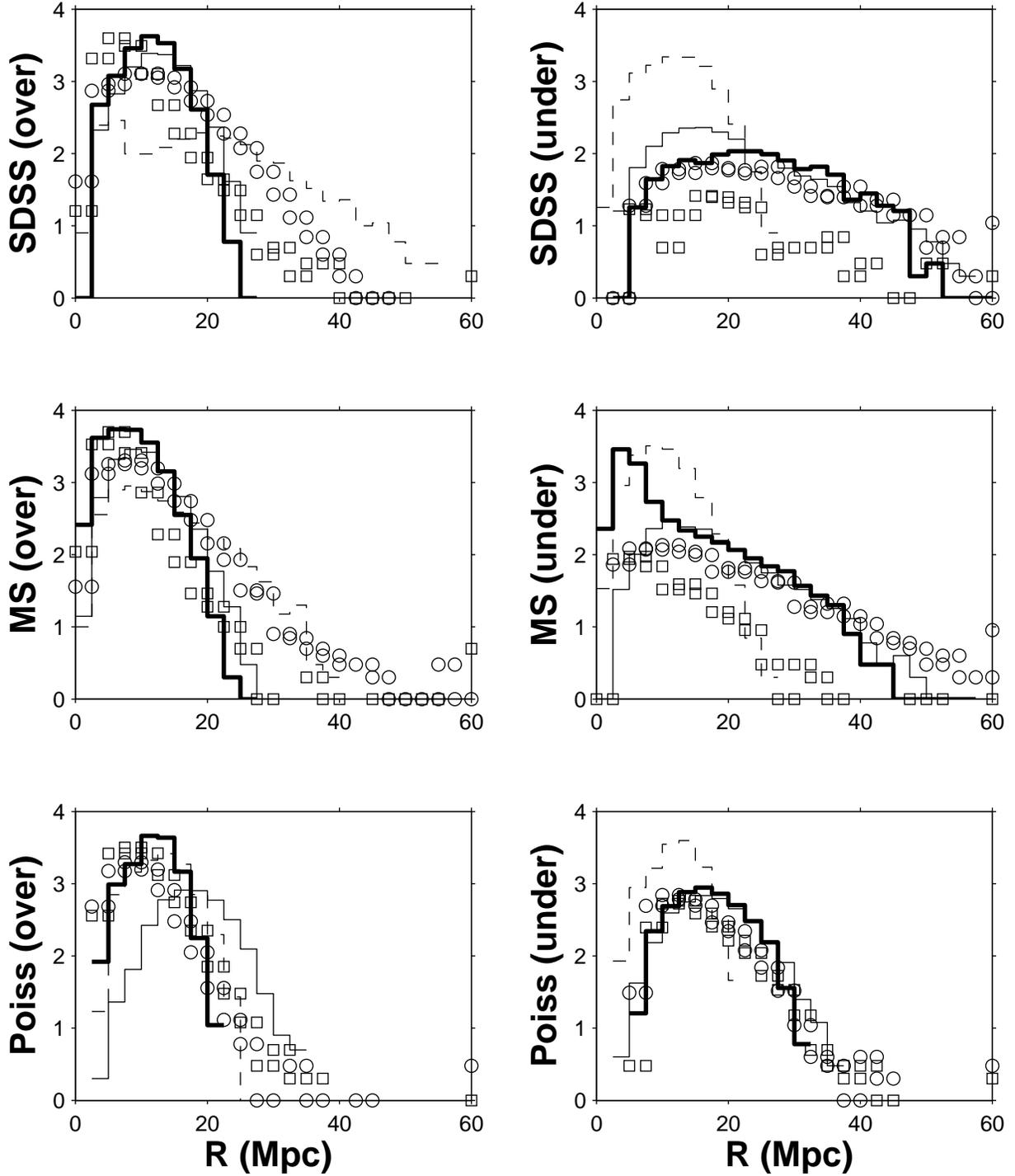}
\caption{Distributions of the effective radii of groups connected with density
maxima (left) and density minima (right) found with HOP for the three standard
data samples: SDSS (top), MS (middle), and Poisson (bottom).  Thick Solid:
Voronoi; solid: KDE; dashed: SOM; circles: BB (volume) and squares: BB (density). 
In each case the base-10 log of the number distribution is plotted against
the effective radius in Mpc.}\label{figure04}
\end{figure}

For a representative selection of the cases in Tables \ref{table01},
\ref{table03}, and \ref{table04} Figure \ref{figure04} plots normalized
distributions of the structures' effective radii, defined in terms of its volume
$V$ by
\begin{equation}
R_{\mbox{eff}} = H_{0} \  ( 3V / 4\pi )^{1/3} \ ,
\label{eqn1}
\end{equation}

\noindent
Here for $V$ we use the sum of the volumes of the Delaunay tetrahedra in the
structure, but alternatively one could use the equal or slightly larger volume
of the convex hull.  The distributions obtained with direct local density
estimates (BB, Voronoi and KDE) are similar, with broad peaks in the range
10-20 Mpc.  The SOM distributions are based on a HOP function that is discrete 
and only indirectly expresses density, so it is not surprising that they are
rather different from the others.  These distributions are quite similar to
that shown in Figure 2 of \cite{pan}.  

\clearpage

\begin{figure}[htb]
\ifcolor
    \includegraphics[scale=1]{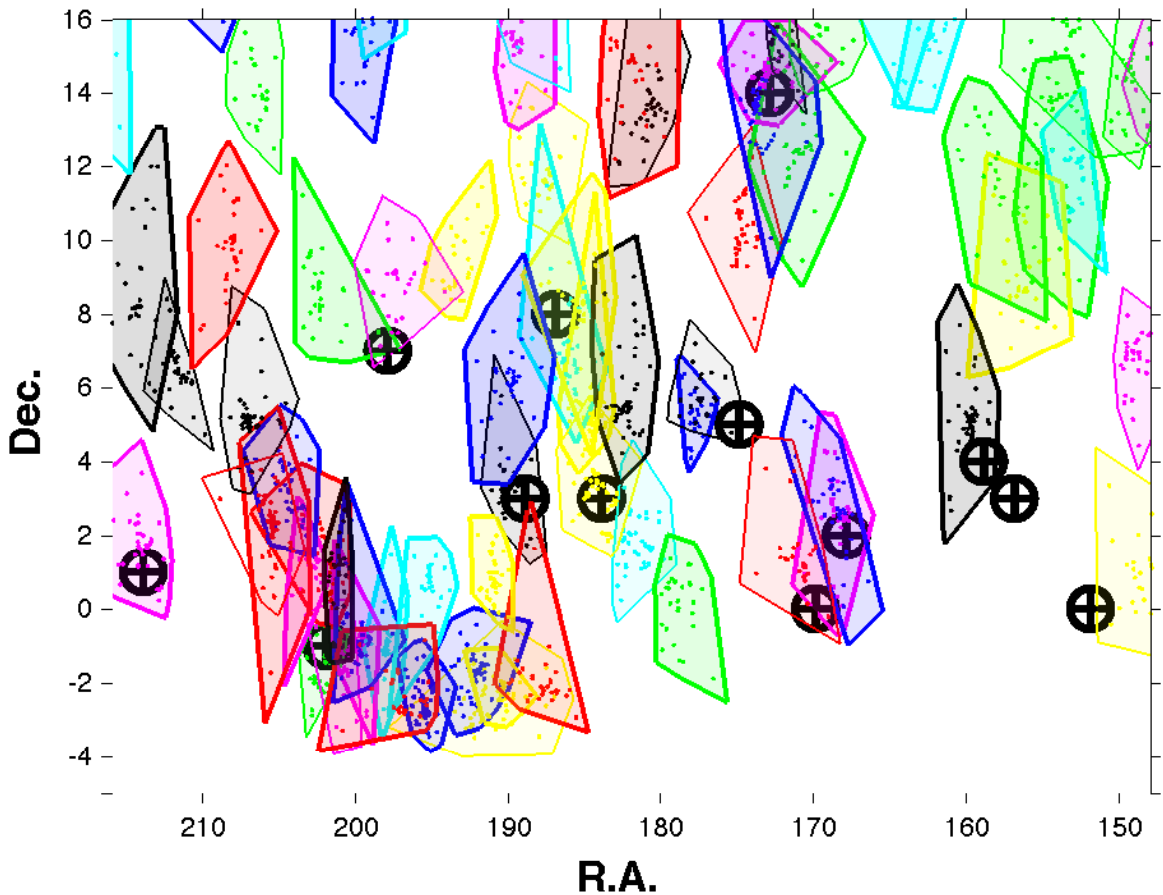}
\else
   \includegraphics[scale=1]{figure05bw.eps}
\fi
\caption{Sky distribution of galaxies for the region of the Sloan Great Wall.  
The HOP groups are delineated by 
\ifcolor
polygons filled with random colors
\else
lightly shaded polygons 
\fi
These are the projected 2D convex hulls of the sky positions of galaxies
contained in the group, slightly expanded to improve the visualization.
The opacities of the polygons are linear in the redshift for the group: 
close darker ones thus appearing to be in front of more distant lighter ones.
The
\ifcolor
heavy black
\else
dark
\fi
circles with $+$ signs are nominal positions of the 13 superclusters in this
region given in Table 1 of \cite{einasto_gw}.}
\label{figure05}
\end{figure}

The region of the Sloan Great Wall is perhaps the richest region of the nearby
Universe.  Figure \ref{figure05} compares our group structures from Voronoi cells
alone (corresponding to row 4 of Table \ref{table02} labeled $1/V_{Voronoi}$;
\ifcolor
   colored polygons) 
\else
   lightly shaded polygons)
\fi
with superclusters in this region (crosses inside circles). The bulk of the Sloan
Great Wall is in the lower-left quadrant.  This figure is limited to galaxies in
the redshift range .045 -- .085 ascribed to the Great Wall.  To eliminate some
clutter only HOP groups with 25 or more galaxies and projected areas of more
than 16 square degrees are shown.  The correspondence with previously cataloged
superclusters is not one-to-one, as expected because of the very different
detection principles involved.  


\subsection{Some Properties of Delaunay Voids}\label{empty}

We now discuss voids in more detail, pointing out some potential problems that
need to be addressed, including the possibility of galaxies lying inside 
Delaunay voids and edges effects similar to those mentioned above and in Paper I
for dense structures.

The Delaunay tetrahedron method described above yields volumes almost completely
devoid of galaxies in the sample. From the way these void structures are
constructed the galaxies at the vertices of the component tetrahedra 
for the most part lie on the void surface leaving the inside empty.
A surface galaxy is one from which there is a
path to the outside that does not intersect any of the tetrahedra.
With this definition a galaxy is inside a Delaunay void if and only if 
all of the void's tetrahedra that have it as a vertex 
cover the full solid angle ($4\pi $ steradians) as seen from that galaxy.
A simple but effective procedure to identify Delaunay voids and 
identify possible interior galaxies is as follows:

\begin{itemize}
\item Compute the Delaunay tessellation of the galaxy positions
\item Identify groups of tetrahedra making up voids using HOP with $\bf{f} =$
tetrahedral volume
\item For each such Delaunay void, containing $N_{void}$ tetrahedra:
\begin{itemize}
\item Collect a list of all $4 N_{void}$ triangular faces of the tetrahedra
making up the void
\item Identify the faces that appear in this list only once.
\item The vertices of such faces are on the surface.
\end{itemize}
\item Identify as internal galaxies any that are not on the surface 
\end{itemize}

In summary we define surface galaxies as those that populate the hull (not to be
confused with the convex hull) of the galaxies circumscribing the void; any void
galaxies not on the hull are then internal. In our analyses this is not an issue:
\emph{not a single one of the many HOP-found Delaunay voids reported here 
contains any internal galaxies}.  Clearly our analysis method militates strongly
against such cases, but we do not know if they are impossible or just extremely
rare.

\begin{figure}[htb]
\includegraphics[scale= 1]{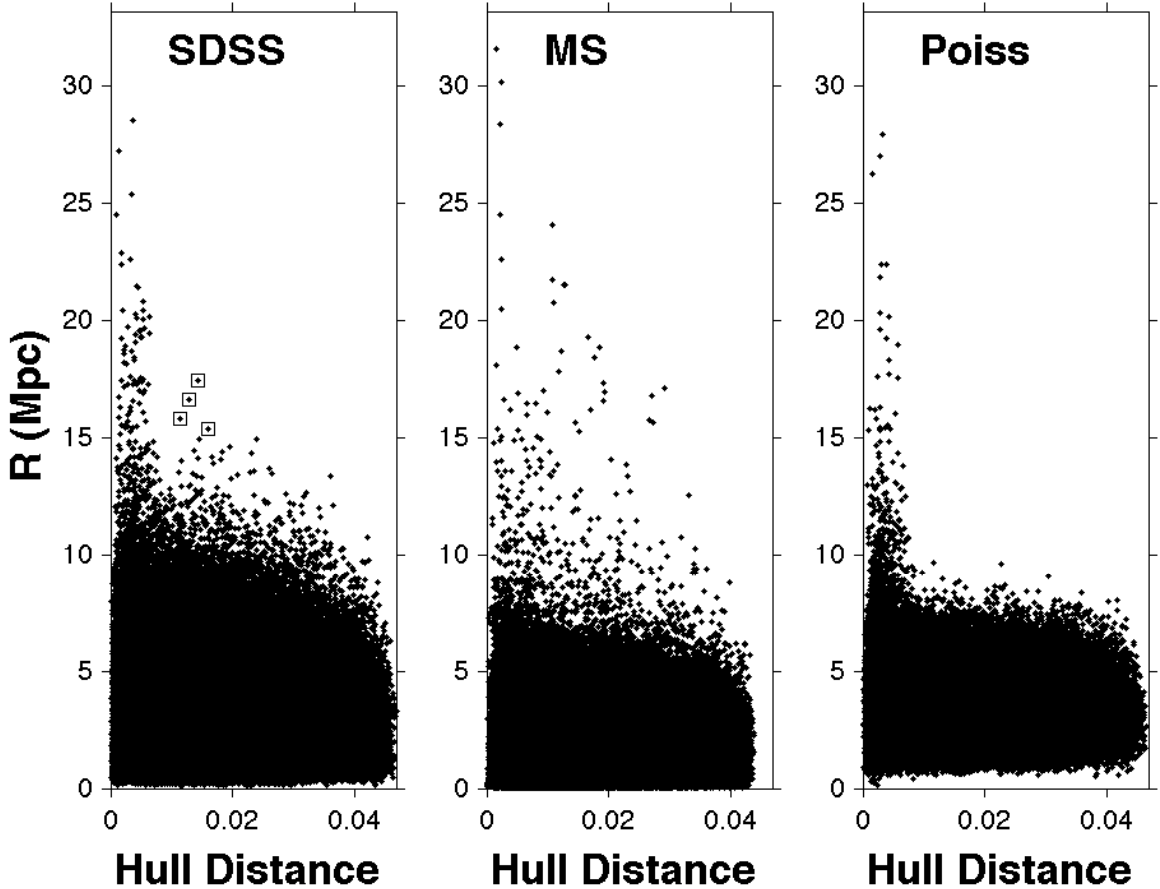}
\caption{The effective radius of each Delaunay void 
is plotted against the minimum distance
from its circumscribing galaxies to vertices
of the augmented convex hull of the full data set. 
In all 3 cases (SDSS: left, MS: middle, Poiss: right)
the sprinkling of large-radius voids at small distances
from the hull are artificially enlarged due to edge effects. }
\label{figure06}
\end{figure}

While edge effects in Delaunay tessellations are less serious than in Voronoi
tessellations, a second potential problem is that tetrahedra at the edges of
the data space are systematically larger than they would be if not located there.
Due to the complexity of the SDSS boundaries in three dimensions
an automated test for whether or not a tetrahedron is at or near an edge
is difficult.  Here we compute the minimum distance of the four galaxies in a
tetrahedron from the nearest point on the convex hull of the data.
A complication arises when the outward-facing triangles
of tetrahedra at the edge are exceptionally large, for then
this minimum inter-galaxy distance is not actually representative
of the distance from the edge. This difficulty is easily circumvented by adding 
points just outside these triangles, thus creating an augmented convex hull,
faithful to the actual one but with no large faces.
Figure \ref{figure06} gives scatter plots of 
effective radius vs. distance from the augmented hull,
clearly showing this inflation effect for voids
within approximately .005 redshift units of the hull.


\begin{figure}[htb]
\ifcolor
\includegraphics[scale= .8]{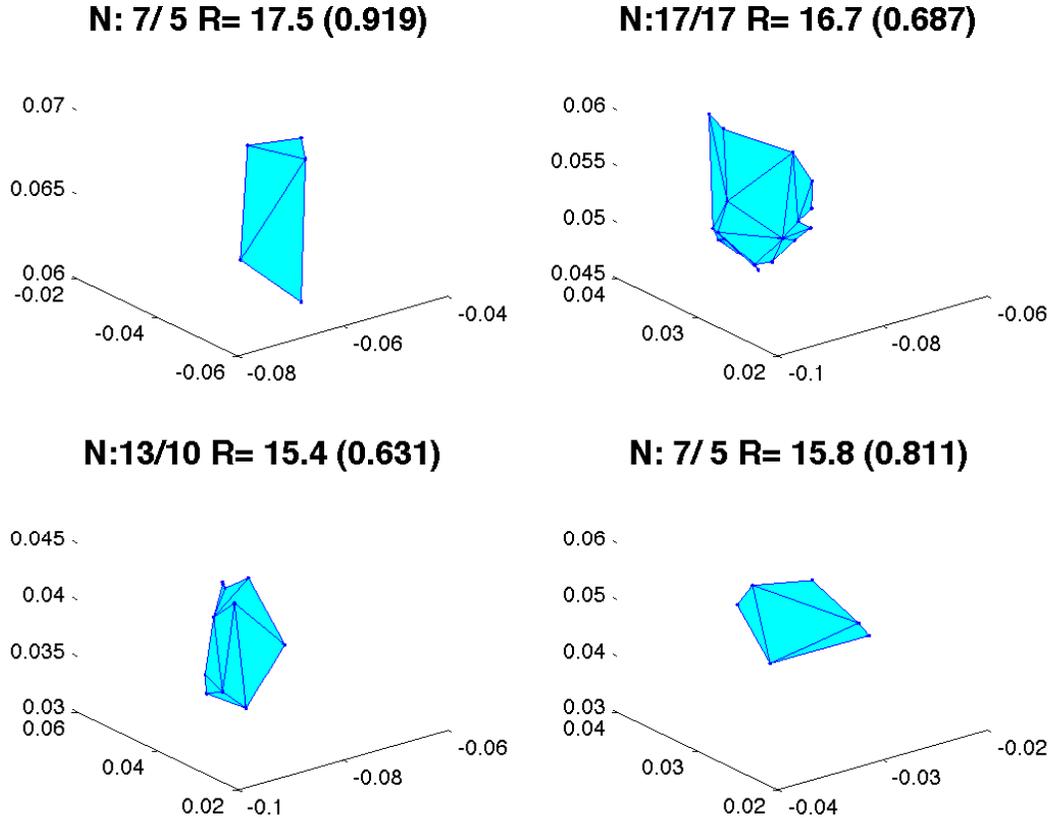}
\else
\includegraphics[scale= .8]{figure07bw.eps}
\fi
\caption{The four largest SDSS Delaunay voids judged to be free of edge effects, 
and indicated with squares in the upper-left panel of Fig. \ref{figure06}. 
The two integers in the  legends are the number of circumscribing galaxies and 
tetrahedra, respectively -- followed by the effective radius in Mpc and in
parentheses the convexity (the sum of volumes of the tetrahedra divided by the
volume of the convex hull of the galaxies).}\label{figure07}
\end{figure}

Figure \ref{figure07} shows 3D plots of the four largest SDSS voids that are
farther than .007 redshift units from the augmented hull (indicated by squares 
drawn around the points in Fig. \ref{figure06} above).  The largest of these
voids (upper left panel) is centered at
$RA = 14.5 \pm 0.2 ^{h}, DEC = 39.4 \pm 2.2^{o}$, and 
$z= 0.1041 \pm .0034$, and is therefore near and possibly associated with the
so-called \emph{Bo{\"o}tes} or \emph{giant void}, given various positions by
different authors  -- e.g.  RA $\approx 13^{h}$ [11.5 - 14.3],
DEC $\approx 40^{o}$  [26.5 - 52.0], and z  $\approx$ 0.11 by \citet{kopy}.
Theirs is a very different kind of structure, an order of magnitude larger in
linear size and containing, according to these authors, not just galaxies but 
17 clusters in the ranges shown. It is clear that we are finding very different
void structures than those obtained with other methods.

Voids are typically considered to be regions where the density of galaxies is
$\approx 10\%$ of the mean density and 10's of megaparsecs in size
\citep[e.g.][]{pan,coil,patiri}.  By definition the Delaunay voids derived here
have no galaxies within the analyzed data set and hence have close to $0\%$ of
the mean density.

\begin{figure}[htb]
\includegraphics[scale=1]{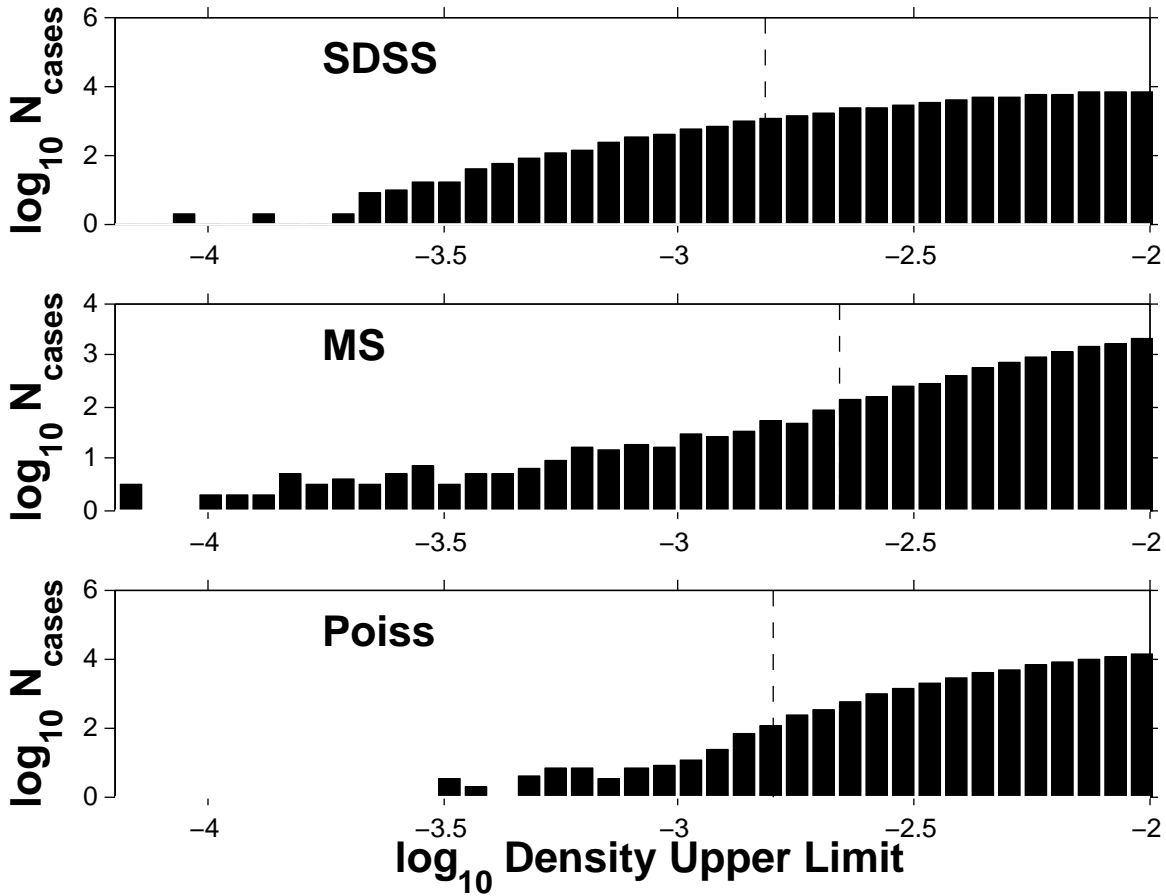}
\caption{Distribution of upper limits (galaxies per cubic Mpc) implied by the
emptiness of Delaunay voids.}\label{figure08}
\end{figure}

Figure \ref{figure08} gives some insight on the information about densities
implied by our empty voids.  If the number of galaxies in a given volume 
follows the Poisson distribution, the likelihood $P(\lambda) = e^{- \lambda V}$
for the rate $\lambda$ (galaxies per unit volume) in an empty volume $V$ gives an
upper limit of $\lambda_{ul} = - \mbox{log}( p_0 ) / V$ with a confidence of
$1-p_0$.  The figure plots the distribution of this quantity for $p_0 = 0.05$
(equivalent to a 95\% confidence).  In order to avoid cells with inflated volumes
due to edge effects a cut of .0055 redshift units (205 Mpc) was applied on the
minimum distance between the void vertices and the nearest face of the convex
hull of the full data set.  In addition most of the large number of upper limits
larger than the overall mean density (shown as a vertical dashed lines) 
are simply not shown.  These are not rigorous upper limits, and the only point is
that the SDSS and MS samples yields a number of voids with lower limits
considerably smaller than 10 percent of the mean density and 
somewhat lower than the random sample.
Further density information about the voids detected may come from a census of
SDSS galaxies not included in our volume limited (VL) sample. Some of those not
included in the VL sample will contain spectroscopic redshifts to the limit
of the Main-Like galaxy sample (m$_{r}<$18) (cf. Section \ref{subsec:stage2})
while others may contain photometric redshifts to the limit of the photometric
sample (m$_{r}<$21) \citep[see,][]{York2000,Strauss02}.  A future paper will
explore density limits in voids, as well as other descriptors of structures.

\section{Conclusions}\label{conclusion}

We have provided tools and data products to explore the multi-scale structure of
the distribution of galaxies, using the SDSS DR7 redshift survey data, 
and in comparison to simulated and random data. The procedure demonstrated here 
departs from other work in use in several ways. Starting with structural
building blocks in the form of tessellation elements or small collections of them,
we use the HOP algorithm to identify structures connected with density maxima
and minima, using the density estimators described in Paper I for the HOP
function, coupled with adjacencies defined by Voronoi-Delaunay tessellation.
Our HOP-based procedure is much simpler than those based directly on discrete
Morse theory.  Nevertheless it identifies all local density maxima and minima 
plus their descending and ascending manifolds.  Furthermore we eschew methods
such as topological persistence because, by eliminating structures based on a
notion of importance, they unnecessarily discard valuable information.
Such methods may allow one to concentrate on certain salient features, 
but the results are then dependent on the choices of \emph{importance quantifier}
and methods for discarding, combining or otherwise modifying features.

Note that Nature does not single out any one definition of structural elements or
procedures for identifying and characterizing them.  Methods invoking other than
the purely geometrical information utilized here, such as colors or gravitational
binding, undoubtedly yield very different structural descriptions. This
dependence on methodology is not an uncertainty, statistical or otherwise,
but an inevitable and useful feature of the diversity of analysis approaches.

A few summary statistics are presented here and further details will be presented
in future publications.  Three dimensional distributions of multi-scale structure
are not easily displayed in a paper.  We encourage the reader to explore
visualizations of the data given as electronic-only material.  Such displays
might be compared to the density map shown in Figure 2 of \citet{gott}
which suggests a visual similarity of the distribution of the highest 7\% and
lowest 7\% of a smoothed and pixelated density distribution.  Such displays of
course cannot convey a complete visualization, rather they show that these two
spatial distributions of very different quantities (one of galactic density,
the other of degree of local sparsity) are at least superficially quite similar.
A future paper will explore this similarity by investigating auto-correlation
functions, cross-correlation functions, and other statistical techniques,
and make detailed comparison with similar collections of multi-scale results
such as dense structures  \citep{park_great_wall} and voids \citep{sutter}.

All of the data needed to construct structure catalogs are contained in
electronic-only files described fully in Appendix A.  We provide files containing
information to construct catalogs of multi-scale density and void structures
based on several modes of analysis of the data, with a special eye toward
comparing SDSS structures with those in simulation and purely random data sets.
Each galaxy is assigned to a structure defined by a local maxima, and the
collection of these for a given maximum define a max-structure (also called a
group or cluster).  In addition each galaxy is also assigned to a structure
defined by a local minimum, with the collections defining min-structures
(or voids). One can apply cutoffs in order to limit the outskirts of individual
structures, procedures to eliminate structures that are not significant according
to a given criterion, and possibly other post processing techniques.

The geometrical structures in the spatial distribution of galaxies, here termed
\emph{the multi-scale structure of the Universe}, have been said to form a
\emph{hierarchy} \citep{weygaert2009,neyrinck2008,aragon_calvo_3,aragon_calvo_4},
often without a precise definition of what this means.  This nomenclature implies 
the existence of a set of \emph{discrete} hierarchically related levels.
These ideas seem to be derived from concepts such as merger trees in theoretical
or phenomenological models of structure formation; cf. the discussion in
\cite{knebe}.  It is possible that signatures of such effects can be detected in
the current multi-scale structure, say in the form of clusters of clusters of
galaxies (popularly referred to as superclusters) and the like
\citep[but see][]{yu_peebles}.
However studies of characteristics of multi-scale structure in the
universe, including the present work, demonstrate only continuous distributions
and fail to show evidence of discreteness in either qualitative characteristics
or quantitative observables.  A continuous or self-similar distribution
\citep[e.g.][]{einasto89} is if anything the opposite of a discrete hierarchy.
A similar point has been made by \cite{peebles_hierarchy,peebles_3}.
In short we see no evidence of a discrete hierarchy in the multi-scale structure
of the Universe.  Note that the discrete density levels of Bayesian Blocks or
Kernel Density Estimation described here and in Paper I are a contrivance for
density representation and have nothing to do with a discrete hierarchy of any
kind in the actual distributions.

The catalogs and other data products given here can be utilized by any group to
compare the structures found by any technique and make them immediately
comparable to those of another.  Future papers in the current series will 
describe more detailed statistical summaries of geometric and topological
properties of both over-dense, under-dense, and empty structures,
and carry out various comparisons with previous cluster and void catalogs.

\acknowledgements
We are grateful to the NASA-Ames Director's Discretionary Fund
and to Joe Bredekamp and the NASA Applied Information
Systems Research Program for support and encouragement.
Thanks goes to Ani Thakar and Maria Nieto-Santisteban for
their help with our many SDSS casjobs queries. Michael Blanton's
help with using his SDSS NYU--VAGC catalog is also very much appreciated.
We are grateful to Patrick Moran, Christopher Henze, Changbom Park,
Paul Sutter, Mark Neyrinck, Thierry Sousbie, Tom Abel, Pratyush Pranav,
Peer-Timo Bremer, Attila Gyulassy, James (``B.J.'') Bjorken and Jessi Cisewski.
Special thanks goes to Slobodan Simic and members of the CAMCOS project at San
Jose State University, Joseph Fitch, David Goulette, Jian-Long Liu,
Mathew Litrus, Brandon Morrison, Hai Nguyen Au, and Catherine (Boersma) Parayil
for useful comments and for an ongoing collaboration on developments of the
HOP algorithm for topological data analysis.  None of these acknowledgements
should be construed to imply agreement with the ideas expressed here.

Funding for the SDSS has been provided by
the Alfred P. Sloan Foundation, the Participating Institutions, the National
Aeronautics and Space Administration, the National Science Foundation,
the U.S. Department of Energy, the Japanese Monbukagakusho, and the Max
Planck Society. The SDSS Web site is http://www.sdss.org/.

The SDSS is managed by the Astrophysical Research Consortium for
the Participating Institutions. The Participating Institutions are The
University of Chicago, Fermilab, the Institute for Advanced Study, the
Japan Participation Group, The Johns Hopkins University, Los Alamos National
Laboratory, the Max-Planck-Institute for Astronomy, the
Max-Planck-Institute for Astrophysics, New Mexico State University,
University of Pittsburgh, Princeton University, the United States Naval
Observatory, and the University of Washington.

This research has made use of NASA's Astrophysics Data System Bibliographic
Services.
This research has also utilized the viewpoints \citep{GLW2010} software package.
\clearpage

\section{Appendix A: Data Details}
\label{appendix_a}

Of the three data sets studied, the first is a volume limited sample of 146,112
galaxies drawn from the SDSS.  The second catalog is drawn as similarly as
possible from the MS, yielding a volume limited sample of 171,388 galaxies.
Third is a set of 144,700 points mimicking the SDSS volume limited sample 
but randomly and independently distributed so that there is no 
spatial structure beyond that imposed by the SDSS sampling.\footnote{Such 
identically and independently distributed (IID) processes are often
called Poisson processes (here with a spatially constant event rate) 
because the counts in fixed volumes obey the Poisson distribution.}
All conversions from redshift coordinates to Mpc are based on a Hubble
constant of $73$ km s$^{-1}$ Mpc$^{1}$.

\subsection{The SDSS NASA/AMES Value Added Galaxy Catalog (AMES--VAGC)}\label{sec:sdsscatalog}
This section provides details in addition to those given in Paper I.
The NASA/Ames Research Center SDSS Value Added Catalog (NASA--AMES--VAGC)
is based on the New York University Value Added Catalog
\citep[NYU--VAGC][]{Blanton05}, that is in turn derived from Data Release 7 of
the SDSS \citep{SDSS7}.  We now describe the stages in the catalog creation.

\subsection{Stage 1: Extracting tables from the SDSS NYU--VAGC}\label{subsec:stage1}

The contents of a number of NYU--VAGC fits table files (described below) were 
extracted and used to create {\it Stage 1} of the catalog.  An index of those
fits files is listed below.  At the time the catalog was created only the
NYU--VAGC had SDSS K-corrected absolute magnitudes readily available and hence
we did not originally use the catalogs available via the excellent SDSS casjobs
server.\footnote{http://casjobs.sdss.org}

Selections were applied to each of the following three NYU--VAGC fits files:

\begin{itemize}
\item object\_sdss\_spectro.fits:
	\begin{itemize}
	\item SDSS\_SPECTRO\_TAG: Galaxy Spectrum exists
	\item PRIMTARGET: Select Main Galaxy Sample targets
	\item OBJTYPE: Select type GALAXY
	\item CLASS: Select type GALAXY
	\item Z: Estimated redshift
	\item Z\_ERR: Estimated redshift error. Only allowed to be greater than
		zero since negative values indicated an invalid estimate
	\item ZWARNING: Must be equal to zero to indicate no warning flags
		in the redshift estimation procedures
	\end{itemize}
\item object\_sdss\_imaging.fits:
	\begin{itemize}
	\item RA: Right Ascension
	\item DEC: Declination
	\item NCHILD: Must be zero indicating that it is not part of a
		blended parent or blended itself (!BLENDED)
	\item RESOLVE\_STATUS: Used to obtain only one instance of each object
	\item VAGC\_SELECT: To satisfy the Main-like criteria of the NYU--VAGC
	\item FLAGS: Include only !BRIGHT, !BLENDED, !SATURATED
	\item MODELFLUX: Model Magnitude fluxes (extinction corrected)
	\item MODELFLUX\_IVAR: Inverse variance of the fluxes (flux errors)
	\item PETROR50: 50\% Petrosian Radius
	\item PETROR90: 90\% Petrosian Radius
	\end{itemize}
\item kcorrect.none.model.z0.10.fits:
	\begin{itemize}
	\item ABSMAGS: Absolute magnitudes in U,G,R,I,Z,J,H,K using
	a 0.1 blue-shift of the band-passes for k-corrections
	\end{itemize}
\end{itemize}

\begin{figure}[htb]
\includegraphics[scale=0.9]{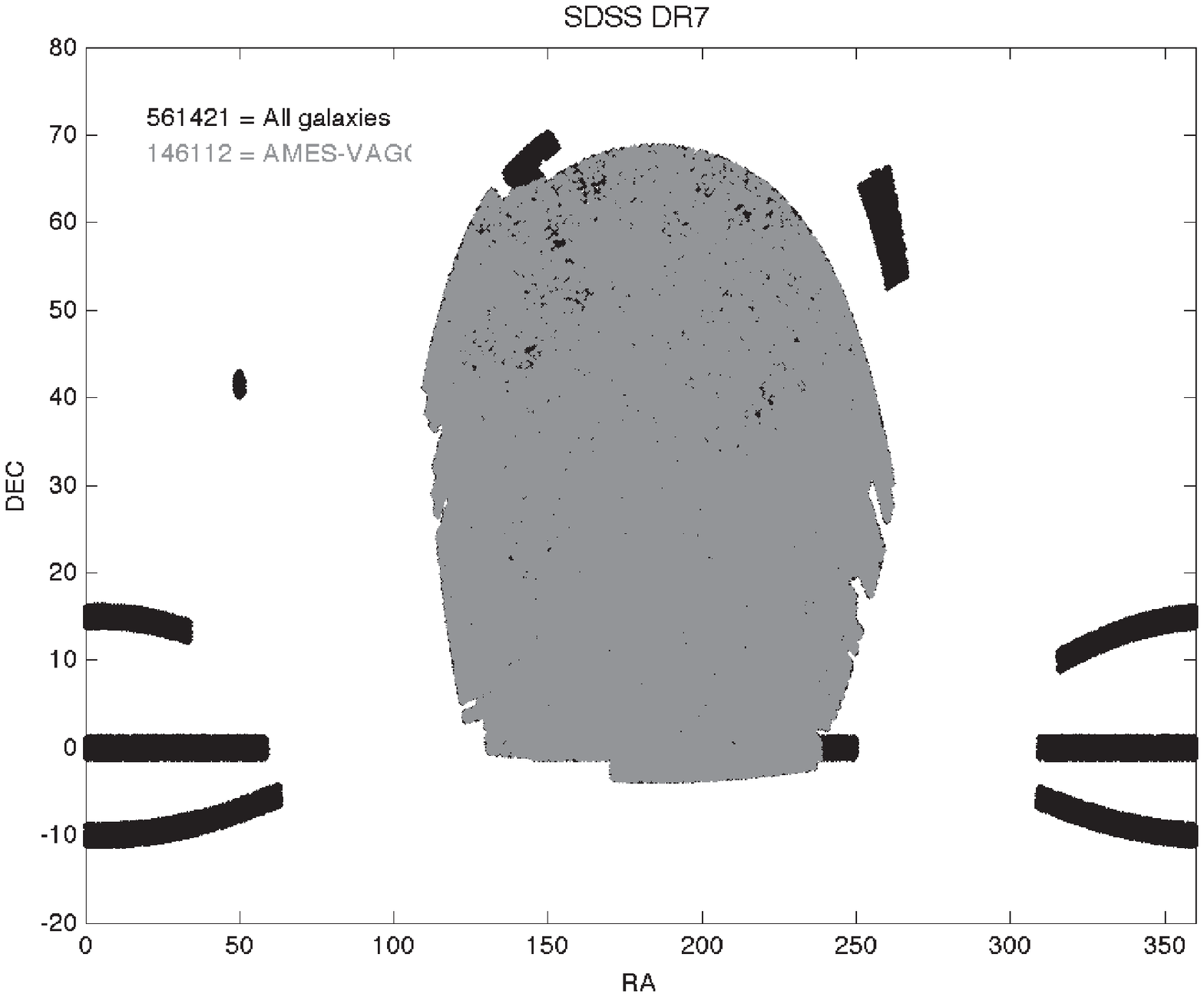}
\caption{Plot of the entire SDSS DR7 Main-Like Galaxy Sample from the NYU--VAGC
catalog (both black and gray points).  The points in gray are those of the
volume limited sub-sample derived from stage two of the catalog as described in
Section~\ref{sec:sdsscatalog}. The black points were eliminated from the volume
limited sample as a result of redshift and absolute luminosity cuts,
($z <$0.12 and M$_{R}^{0.1}<$--20.0751), and the desire for a contiguous
geometric sample.} \label{figure09}
\end{figure}

The outputs of these selections were concatenated into a single {\it Stage 1}
NYU--VAGC Main--Like Galaxy Sample catalog containing 561,421 galaxies. See
Figure~\ref{figure09} for a plot of the points in Right Ascension
(RA) and Declination (DEC). The catalog at this
stage contained an internally assigned identification number, 
RA, DEC, apparent magnitudes (u,g,r,i,z), apparent magnitude errors,
absolute magnitudes (U,G,R,I,Z,J,H,K), absolute magnitude errors,
redshift, redshift error, Petrosian 50\% and 90\% radii.

\subsection{Stage 2: Obtaining a contiguous and volume limited sample}\label{subsec:stage2}

The maximum number of galaxies in our volume limited sample consistent with
common practices in using the SDSS turned out to be 163,157.  
These selections \citep[e.g.][]{Choi2010} are redshift $z<$0.12, absolute
magnitude in the r bandpass M$_{R}^{0.1}<$20.0751. These are consistent with a
red band apparent magnitude upper limit defined by the \cite{Strauss02} Main
Galaxy Sample as r$<$17.77, although the NYU-VAGC Main-like sample goes down
to $r = 18$.

Next samples were removed outside a defined contiguous region avoiding several
irregular features extending beyond the smooth outer (two dimensional) shape of
the distribution of points, as well as disconnected and isolated patches lying
entirely outside.  This region was centered on the north galactic cap roughly
corresponding to 100$<$RA$<$270 and -7$<$DEC$<$65.  The contiguous region
contains 146,112 objects and is defined by the gray area in
Figure~\ref{figure09}.

\subsection{Stage 3: 55$\arcsec$ fiber placement issue and coordinate transform}\label{subsec:stage3}
The angular separation in arc--seconds to the 6 nearest neighbors for
every point was estimated. This allows one to quickly identify any neighbor
within 55$\arcsec$.  This was necessary because the fiber plug plate
of the SDSS does not allow fibers to be placed closer than 55$\arcsec$
to each other. However, there are a large number of overlapping plates
which means that there are some galaxies with spectra within this
55$\arcsec$ fiber limit.  Since these overlaps cover only part of the
full area it represents a systematic bias that must be eliminated in order
to consistently sample the true underlying galaxy distribution.
To do so we removed a randomly chosen member of any pair found within 55$\arcsec$
of each other.  This process eliminates 6,314 galaxies from the
sample.\footnote{In Paper I it was claimed that 6,540 galaxies were eliminated,
but this is incorrect.}

To use Euclidean coordinates with units the same in all 3 dimensions the 
right ascension ($\alpha$), declination ($\delta$), and redshift ($z$) were 
transformed into Cartesian coordinates according to 
\begin{eqnarray}
 & \mbox{x}& \mbox{ = z cos}(\delta) \ \mbox{cos}(\alpha)  \label{radec2xyz_1}\\
 & \mbox{y}& \mbox{ = z cos}(\delta) \  \mbox{sin}(\alpha) \label{radec2xyz_2} \\
 & \mbox{z}& \mbox{ = z sin}(\delta) \label{radec2xyz_3}
\end{eqnarray}

\noindent (equivalent to the MatLab\copyright \ function \verb+sph2cart+)
thus yielding rectangular coordinates, each with units of redshift and
convertible to physical units by multiplying by $c /H_{0}$, with $c$ the speed
of light and $H_{0}$ the Hubble constant.  A nonlinear conversion can also be
made for a given cosmological model, but will yield only a small correction 
over the low redshift range of these data.

\subsection{Stage 4: Voronoi related calculations}\label{subsec:stage4}

The Voronoi tessellation \citep[e.g.][]{spatial_tessellations} of the remaining
galaxies was calculated (see Paper I for more details).  From this tessellation 
a number of additional parameters are derived:

\begin{enumerate}
\item Cell volume: V
\item The distance between each galaxy and the center of its Voronoi cell: $d_{CM}$
\item The minimum and maximum dimension of each Voronoi cell: R$_{min}$, R$_{max}$
\item Cell radius: R$_{Voronoi}=({3V \over 4\pi})^{1/3}$
\item A measure of cell elongation: E=$R_{min} \over R_{max}$
\item A measure of the magnitude of the local density gradient: $d_{CM}$/R$_{Voronoi}$
\item A scaling parameter for distances: the average density of the volume
limited SDSS data raised to the minus 1/3 power:
d$_{uniform}$=3.2$\times$10$^{-3}$ in units of redshift\label{scale_factor}
\end{enumerate}

\noindent The first three are fundamental properties of the Voronoi cells.
They are defined for individual cells but are dependent on neighboring galaxies
by virtue of the way the Voronoi tessellation is defined. In turn they are
used to derive useful properties 4, 5 and 6. The first two of these are summary
descriptions of the size and shape of the cell. The separation between 
each galaxy and the center of its Voronoi cell is a vector that approximates 
the magnitude and direction of the local gradient in the density of galaxies.
It is here represented by its magnitude in item 6.

The average distance in item 7, a property of the full sets of galaxies in the
catalogs, is not used in the assignment of individual galaxies to classes.  
Instead, it is used as a scaling factor to make distance parameters such
as $d_{CM}$ and $R_{Voronoi}$ dimensionless.  The average distance here
is computed as the average spacing, $(V/N)^{1/3}$, between samples.
The actual value for the Millennium simulation was very near that of the SDSS,
while the Random was set to this value when the data set was created.  This
quantity was chosen because it is well--defined, straightforward to calculate,
and insensitive to details such as the usage of the Voronoi tessellation algorithm.

\subsection{Stage 5: Flagging boundary points}\label{subsec:stage5}

The cells near the boundaries of the tessellated volume are distorted to one
degree or another.  Depending on the distance of the cell from the boundary, 
this effect ranges in importance from small to large.
The most distortion happens when the tessellation algorithm 
assigns to a cell one or more vertices well outside the data volume, or 
even leaves a vertex undefined because it formally lies at infinity.  
One could attempt to correct for such distortion but
as described in Paper I we feel it is better to simply eliminate 
galaxies whose Voronoi cells appear to have been 
significantly distorted by boundary effects. 
Our criteria for identifying such cells, as detailed in the Section titled
``The Voronoi Cell Boundary Problem'' of Paper I, led to the rejection of 
5807 boundary cells, leaving 133,991 galaxies in the sample
to be used for the SDSS density estimations reported here.

\subsection{Stage 6: Building a table for casjobs}\label{subsec:stage6}

In order to make the sample useful for users of casjobs (where most SDSS
users obtain their data) we have attempted to obtain SDSS object identification
numbers from the PhotoObjAll.ObjID table for all of the objects in the final
density sample.  This was necessary because the NYU--VAGC DR7 catalog does not
contain the same object identification numbers as those found in the SDSS DR7 casjobs
catalog.  To obtain the object identifications the fGetNearestObjAllEq function 
of casjobs was used. 
Objects were matched within 1$\arcsec$ of the RA and DEC
of the NYU--VAGC derived objects.\footnote{Query: select a.*, b.objid as
matchObjID into mydb.nyuvagccross from MyDB.densitycatalog a\newline
cross apply dbo.fGetNearestObjAllEq(a.ra, a.dec, 0.0167) b}
From 146,112 points (see Section 3.2) 145,875 PhotoObjAll.ObjID identifications
were found (known simply as the ObjID in SDSS casjobs parlance),
meaning that 237 points did not exist in the casjobs catalog.  This 0.16\% loss
should not be a major inconvenience for casjobs-based procedures. Those 237
objects in the final NASA--AMES--VAGC catalog without casjobs ObjID numbers will
still be in the publically released catalog, but will instead contain an 18
character string (the same length as the unique SDSS ObjID) with each object
numbered from 000000000000000001 to 000000000000000237.

\subsection{The adaptive kernel map classes}\label{subsec:akmclasses}

In Table 3 of Paper I the Bayesian Blocks (BB) and adaptive kernel map (AKM)
methods had a number of classes that ranged from low density to high.
The class structure for the Self Organizing Map (SOM) method was
more complex (see Table 2 in Paper I). The AKM method produces
a continuous range of densities rather than specific classes. In order
to mimic the BB and SOM class methods a filter was applied to the
AKM densities to produce the 11 classes found in Table 3 of Paper I:
\begin{equation}\label{eq:akm}
AKM_{class}=12-\mbox{round}(((log_{10}(AKM_{density})/5.6947)\times20)-7)
\end{equation}

\begin{figure}[ht!]
\includegraphics[scale=1]{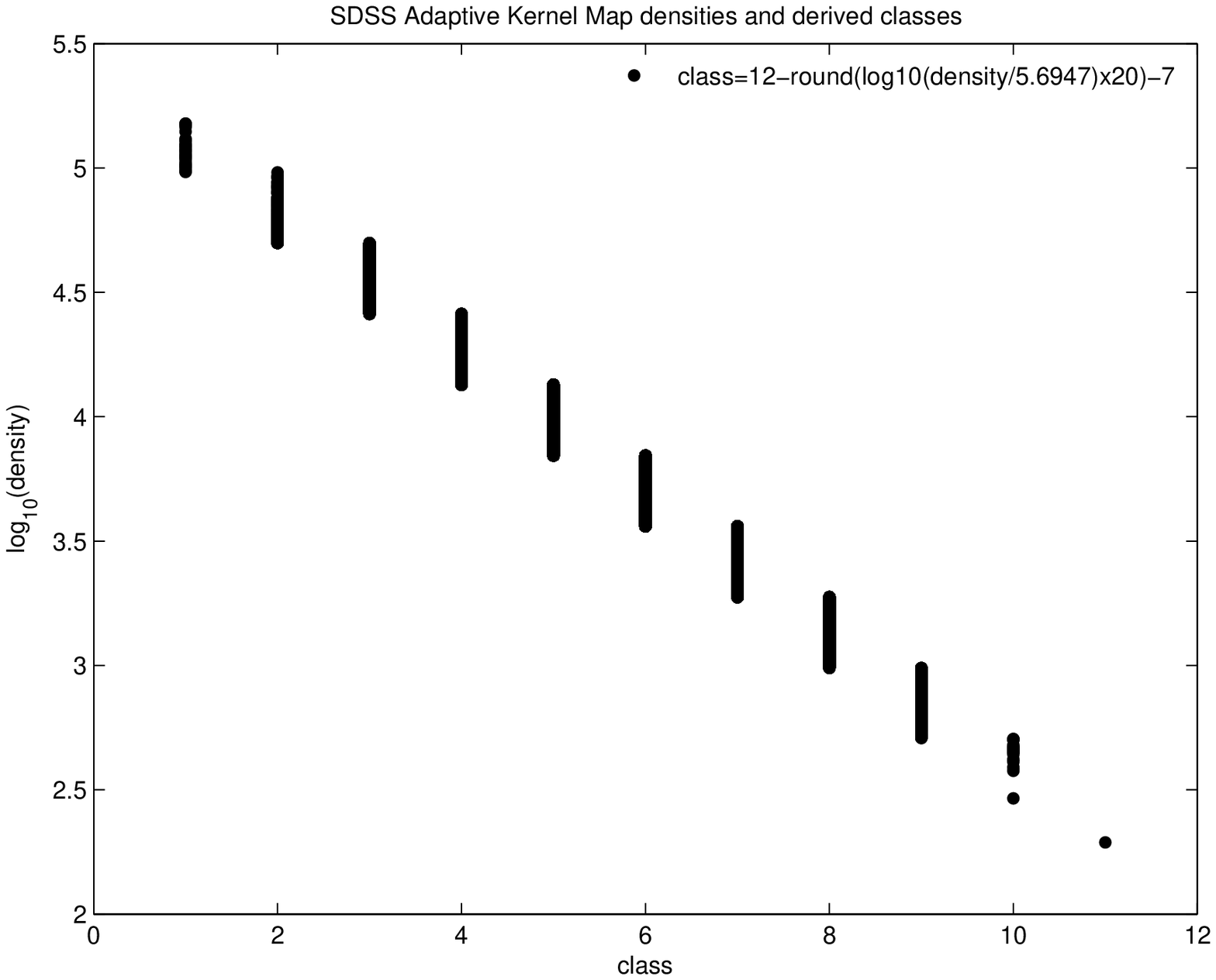}
\caption{Plot of Adaptive Kernel Map densities versus their derived classes
using Eq. \ref{eq:akm} in the text and figure legend.}
\label{figure10}
\end{figure}

Fig. \ref{figure10} shows the resulting correspondence between
AKM density and class.\footnote{A better method to segment the data might have
been to utilize the unique strengths of Bayesian Blocks, but that
was not done herein.}

\subsection{The Millennium Simulation AMES Value Added Catalog}\label{sec:mscatalog}

To create a volume limited sample from the MS a similar procedure was
followed to that described in Section~\ref{sec:sdsscatalog}.
This is possible since one can obtain the absolute magnitude
estimates in the same bandpasses as the SDSS for the galaxies in
the MS \citep{Croton2005}.  First one must convert the MS Cartesian 
coordinates and velocities (x,y,z,v$_{x}$,v$_{y}$,v$_{z}$)
to right ascension, declination, and redshift using 
H$_{o}$=73 km sec$^{-1}$ Mpc$^{-1}$, $\Omega_M$=.25, $\Omega_\Lambda$=.75,
$\Omega_K$=0.  The apparent magnitudes were derived from the given absolute
magnitudes using the luminosity distance. The luminosity distance
requires the redshift and radial distances derived from the MS.\footnote{See
\cite{Cheng2005,Peacock1999,hogg1999} for more on the luminosity distance.}
The same redshift, absolute magnitude cuts as in the AMES--VAGC were applied
leaving 171,388 out of $\sim$9 million points in the original MS catalog.
16,283 points were eliminated to emulate the SDSS 55$\arcsec$ fiber
collision issue, while 6178 were eliminated because of boundary effects. This
leaves 148,927 points.

The same distance scaling factor as used for the SDSS data, as
described in Section \ref{subsec:stage4}, Item \ref{scale_factor},
namely d$_{uniform}=3.2\times10^{-3}$, was used to derive the same Voronoi
quantities found for the SDSS in Section \ref{subsec:stage4}.

Again, in order to mimic the BB and SOM class methods a filter was applied
to the AKM densities to produce the 13 MS classes found in Table 3 of Paper I:
\begin{equation}\label{eq:akm-ms}
AKM_{class}=14-round(((log_{10}(AKM_{density})/5.6947)\times20)-7)
\end{equation}

\subsection{The Randomly Distributed Point Catalog}\label{sec:rand_cat}

The creation of the randomly distributed data point catalog was outlined
in detail in Paper I. The initial data set contains a similar number
of points (144,700) as both the AMES--VAGC and derived MS catalogs.
The final catalog, after removing 6219 points corresponding to the
55$\arcsec$ issue discussed previously and 6649 boundary points discovered
after the Voronoi tessellation yields 131,832 points.\footnote{The final number of
points described in Paper I is incorrect. The 144,700 quoted was before
the removal of the 55$\arcsec$ and boundary value points, not after.}

The galaxy positions were then converted to rectangular Cartesian coordinates
according to the same formulas used for the SDSS data, namely eqs. 
(\ref{radec2xyz_1}), (\ref{radec2xyz_2}) and (\ref{radec2xyz_3}).
As above any transformation that would require picking a value
for the Hubble constant or a cosmological model was avoided.
The same distance scaling factor used for the SDSS and MS data, as
described in Section \ref{subsec:stage4}, Item \ref{scale_factor},
namely d$_{uniform}=3.2\times10^{-3}$, was used to derive the same Voronoi
quantities found for the SDSS in Section \ref{subsec:stage4}.

As in the previous two cases, in order to mimic the BB and SOM class
methods a filter was applied to the AKM densities to produce the 10
uniform classes found in Table 3 of Paper I:
\begin{equation}\label{eq:akm-uniform}
AKM_{class}=11-round(((log_{10}(AKM_{density})/5.6947)\times40)-13)
\end{equation}
\clearpage

\subsection{Structure Catalog Information: Electronic-Only Files}
\label{sec:unicatalog}

For each of the 3 data sets we have constructed a flat ASCII file, the columns of
which contain information, one row for each galaxy, of use for assembling many
kinds of structure catalogs of interest. Many of the entries echo data from the
NYU-VAGC data archive, for the reader's convenience in constructing and exploring 
structure catalogs derived from the new results.  The names of these
electronically accessible files are
\verb+sdss_master.txt+, 
\verb+ms_master.txt+ and 
\verb+poissmaster.txt+,
and the rest of this sections describes their contents 
and a provides a few notes on there use in constructing structure catalogs.
The terms ``Max-structure'' and ``Min-structure'' mean HOP groups 
associated with local maxima and local minima, respectively.

\begin{center}
\begin{longtable}{|l|l|l|}
\caption[Column Identifiers: 146,112 SDSS Galaxies] {\bf{Column Identifiers: 146,112 SDSS Galaxies}} \label{table05} \\

\hline \multicolumn{1}{|c|}{\textbf{ Column Number }} & \multicolumn{1}{c|}{\textbf{ Variable Name }} & \multicolumn{1}{c|}{\textbf{ Description }} \\ \hline 
\endfirsthead

\multicolumn{3}{c}%
{{\bfseries \tablename\ \thetable{} -- continued from previous page}} \\
\hline \multicolumn{1}{|c|}{\textbf{  Column Number  }} &
\multicolumn{1}{c|}{\textbf{  Name  }} &
\multicolumn{1}{c|}{\textbf{   Description   }} \\ \hline 
\endhead

\hline \multicolumn{3}{|r|}{{Continued on next page}} \\ \hline
\endfoot

\hline \hline
\endlastfoot

\rownumber & objid(high)\tablenotemark{a} & NYU-VAGC Identifier \\
\rownumber & objid(low)\tablenotemark{a} & NYU-VAGC Identifier \\
\rownumber & id2  & running index \\
\rownumber & vagc\underscore specobjid  & NYU-VAGC Spectrum ID \\
\rownumber & x & x-Coordinate \\
\rownumber & y & y-Coordinate \\
\rownumber & z & z-Coordinate  \\
\rownumber & ra  & right ascension \\
\rownumber & dec & declination  \\
\rownumber & redshift & observed redshift \\
\rownumber & redshift\underscore err & redshift error \\\hline
\rownumber & u  & apparent u magnitude \\
\rownumber & g  & apparent g magnitude \\
\rownumber & r & apparent r magnitude  \\
\rownumber & i  & apparent i magnitude \\
\rownumber & z & apparent z magnitude  \\\hline
\rownumber & U & absolute u magnitude  \\
\rownumber &  G  & absolute g magnitude \\
\rownumber &  R & absolute r magnitude \\
\rownumber &   I  & absolute i magnitude  \\
\rownumber &  Z & absolute z magnitude  \\
 \rownumber &  J & absolute j magnitude \\
 \rownumber &  H & absolute h magnitude  \\
 \rownumber &  K & absolute k magnitude  \\\hline
 \rownumber &  p50\underscore u & Petrosian 50\% u radius \\
 \rownumber &  p50\underscore g & Petrosian 50\% g radius \\
\rownumber &   p50\underscore r & Petrosian 50\% r radius \\
 \rownumber &  p50\underscore i & Petrosian 50\% i radius \\
 \rownumber &  p50\underscore z & Petrosian 50\% z radius \\\hline
\rownumber &  p90\underscore u & Petrosian 90\% u radius \\
 \rownumber &  p90\underscore g & Petrosian 90\% g radius \\
 \rownumber &  p90\underscore r & Petrosian 90\% r radius \\
 \rownumber &  p90\underscore i  & Petrosian 90\% i radius \\
 \rownumber &  p90\underscore z & Petrosian 90\% z radius  \\\hline
\rownumber &  dCM/R\underscore Voronoi  & centroid  $\rightarrow$ point  (normalized) \\
 \rownumber & R\underscore Voronoi/dUniform\tablenotemark{b} & cell volume / total volume  \\ 
 \rownumber &  R\underscore Max  & Distance from sample to farthest vertex \\ 
\rownumber &  R\underscore Min  &  Distance from sample to nearest vertex  \\ 
\rownumber &  R\underscore Max  / R\underscore Min   &  Elongation \\ 
 \rownumber & cnWinners  & Class ID with most votes  \\
 \rownumber & volume  & Cell volume  \\\hline
\rownumber &  bb\underscore vol\underscore lev & Level ID BB(vol)\tablenotemark{c} \\ 
 \rownumber & bb\underscore vol\underscore blk & Block ID BB(vol)\tablenotemark{c}    \\
\rownumber &  bb\underscore den\underscore lev & Level ID BB(den)\tablenotemark{d}  \\
\rownumber & bb\underscore den\underscore blk  & Block ID BB(den)\tablenotemark{d} \\\hline 
\rownumber &  f55  & 0; but 1 if cell collision test fails \\
\rownumber &  fbad & 0; but 1 if boundary test fails  \\
\rownumber &  density\underscore akm  & KDE density  \\ 
\rownumber &  bandwidth\underscore akm  & KDE bandwidth    \\ 
\rownumber &  levels\underscore akm & KDE density level \\
\rownumber &  ID(Vor,+) )& Max-structure ID; HOP $\bf{f} =$ 1/volume \\
\rownumber &  ID(Vor,-) & Min-structure ID; HOP $\bf{f} =$ 1/volume \\ 
\rownumber &  ID(AKM,+) &  Max-structure ID; HOP $\bf{f} =$ 1/density\underscore akm \\ 
\rownumber &  ID(AKM,-) & Min-structure ID; HOP $\bf{f} =$ 1/density\underscore akm \\ 
\rownumber &  ID(SOM,+) & Max-structure ID; HOP $\bf{f} =$ cnWinners \\ 
\rownumber &  ID(SOM,-) & Min-structure ID; HOP $\bf{f} =$ cnWinners \\ 
\rownumber &  ID(BB(volume),+) &  Max-structure ID; HOP $\bf{f} =$ n(blk)/volume; BB(vol)\tablenotemark{b} \\ 
\rownumber &  ID(BB(volume),-) & Min-structure ID; HOP $\bf{f} =$ n(blk)/volume; BB(vol)\tablenotemark{b} \\ 
\rownumber &  ID(BB(density),+) & Max-structure ID; HOP $\bf{f} =$ n(blk)/volume; BB(den)\tablenotemark{c}   \\ 
\rownumber &  ID(BB(density),-) & Min-structure ID; HOP $\bf{f} =$ n(blk)/volume; BB(den)\tablenotemark{c}  \\ 
\end{longtable}
    \begin{tablenotes}
      \item{(a) }{These long integer identifiers are divided into two parts; 
      the most significant 9 digits (high) and least significant  (low).
      Using Matlab, after executing \verb+load sdss_master.txt+ the string \\
      \verb+[ int2str( sdss_master( :, 1 ) ) int2str( sdss_master( :, 2 )) ]+
      rejoins the two parts.}
    \item{(b) }{dUniform, the average distance between objects in the sample, equals 3.2e-3 redshift units.}
    \item{(c) }{Bayesian Block analysis based on Voronoi cell volume. }
    \item{(d) }{Bayesian Block analysis based on Voronoi cell density. }
    \end{tablenotes}
\end{center}

The last 10 columns of structure IDs  can be used to construct catalogs as
follows.  Let the MatLab variable \verb+index_structures+ denote an array
containing the integers in one of these columns, This contains, for each galaxy G,
the index of the structure to which the galaxy is assigned by the converged
HOP iteration.  Then one can construct an array containing these structure IDs
using the MatLab command

\begin{center}
\verb+ids = unique( index_structures )+
\end{center}

which is also just the array \verb+1, 2, ..., M+ where M is the number of 
structures HOP has identified.  Then for any structure ID \verb+m+ sastifying
\verb+1 <= m <= M+ the indices of the galaxies in that structure (indexed in the
original raw data array, including galaxies that later failed the f55/fbad tests)
can be found from 

\begin{center}
\verb+galaxy_indices = find( index_structures == ids(m) );+
\end{center}

This allows one to compute many things for that structure, such as
the xyz-coordinates of all the galaxies in it, the volume of the structure
(as the sum of the Voronoi volumes), the number of galaxies in it,
and the density in galaxies per unit volume -- using the corresponding
data in the other columns of the master file.  

The following two tables give similar identifications for the MS and Poisson
data files.  Fewer entries have been defined for these data sets,
but the meanings of the parameters which are in common are the same.


\begin{center}
\setcounter{magicrownumbers}{0}
\begin{longtable}{|l|l|l|}
\caption[Column Identifiers:  171,388 MS Galaxies]{\bf{Column Identifiers: 171,388 MS Galaxies}} \label{table06} \\

\hline \multicolumn{1}{|c|}{\textbf{ Column Number }} & \multicolumn{1}{c|}{\textbf{ Variable Name }} & \multicolumn{1}{c|}{\textbf{ Description }} \\ \hline 
\endfirsthead

\multicolumn{3}{c}%
{{\bfseries \tablename\ \thetable{} -- continued from previous page}} \\
\hline \multicolumn{1}{|c|}{\textbf{  Column Number  }} &
\multicolumn{1}{c|}{\textbf{  Name  }} &
\multicolumn{1}{c|}{\textbf{   Description   }} \\ \hline 
\endhead

\hline \multicolumn{3}{|r|}{{Continued on next page}} \\ \hline
\endfoot

\hline \hline
\endlastfoot

\rownumber & objid & NYU-VAGC Identifier \\
\rownumber & id2  & running index \\
\rownumber & x & x-Coordinate \\
\rownumber & y & y-Coordinate \\
\rownumber & z & z-Coordinate  \\
\rownumber & ra  & right ascension \\
\rownumber & dec & declination  \\
\rownumber & redshift & observed redshift \\
\rownumber & u  & apparent u magnitude \\
\rownumber & g  & apparent g magnitude \\
\rownumber & r & apparent r magnitude  \\
\rownumber & i  & apparent i magnitude \\
\rownumber & z & apparent z magnitude  \\\hline
\rownumber & U & absolute u magnitude  \\
\rownumber &  G  & absolute g magnitude \\
\rownumber &  R & absolute r magnitude \\
\rownumber &   I  & absolute i magnitude  \\
\rownumber &  Z & absolute z magnitude  \\
 \rownumber &  dCM/R\underscore Voronoi  & centroid  $\rightarrow$ point  (normalized) \\
 \rownumber & R\underscore Voronoi/dUniform\tablenotemark{a} & cell volume / total volume  \\ 
 \rownumber &  R\underscore Max  & Distance from sample to farthest vertex \\ 
\rownumber &  R\underscore Min  &  Distance from sample to nearest vertex  \\ 
\rownumber &  R\underscore Max  / R\underscore Min   &  Elongation \\ 
 \rownumber & cnWinners  & Class ID with most votes  \\
 \rownumber & volume  & Cell volume  \\\hline
\rownumber &  bb\underscore vol\underscore lev & Level ID BB(vol)\tablenotemark{b} \\ 
 \rownumber & bb\underscore vol\underscore blk & Block ID BB(vol)\tablenotemark{b}    \\
\rownumber &  bb\underscore den\underscore lev & Level ID BB(den)\tablenotemark{c}  \\
\rownumber & bb\underscore den\underscore blk  & Block ID BB(den)\tablenotemark{c} \\\hline 
\rownumber &  f55  & 0; but 1 if cell collision test fails \\
\rownumber &  fbad & 0; but 1 if boundary test fails  \\
\rownumber &  density\underscore akm  & KDE density  \\ 
\rownumber &  bandwidth\underscore akm  & KDE bandwidth    \\ 
\rownumber &  levels\underscore akm & KDE density level \\
 \hline
\rownumber &  ID(Vor,+) )& Max-structure ID; HOP $\bf{f} =$ 1/volumel \\
\rownumber &  ID(Vor,-) & Min-structure ID; HOP $\bf{f} =$ 1/volume \\ 
\rownumber &  ID(AKM,+) & Max-structure ID; HOP $\bf{f} =$ 1/density\underscore akm \\ 
\rownumber &  ID(AKM,-) & Min-structure ID; HOP $\bf{f} =$ 1/density\underscore akm \\ 
\rownumber &  ID(SOM,+) & Max-structure ID; HOP $\bf{f} =$ cnWinners \\ 
\rownumber &  ID(SOM,-) & Min-structure ID; HOP $\bf{f} =$ cnWinners \\ 
\rownumber &  ID(BB(volume),+) & Max-structure ID; HOP $\bf{f} =$ n(blk)/volume; BB(vol)\tablenotemark{b} \\ 
\rownumber &  ID(BB(volume),-) & Min-structure ID; HOP $\bf{f} =$ n(blk)/volume; BB(vol)\tablenotemark{b} \\ 
\rownumber &  ID(BB(density),+) & Max-structure ID; HOP $\bf{f} =$ n(blk)/volume; BB(den)\tablenotemark{c}   \\ 
\rownumber &  ID(BB(density),-) & Min-structure ID; HOP $\bf{f} =$ n(blk)/volume; BB(den)\tablenotemark{c}  \\ 
\end{longtable}
    \begin{tablenotes}
    \item{(a) }{dUniform, the average distance between objects in the sample, equals 3.2e-3 redshift units.}
    \item{(b) }{Bayesian Block analysis based on Voronoi cell volume. }
    \item{(c) }{Bayesian Block analysis based on Voronoi cell density. }
    \end{tablenotes}
\end{center}



\begin{center}
\setcounter{magicrownumbers}{0}
\begin{longtable}{|l|l|l|}
\caption[Column Identifiers: 144,700 Random Points]{\bf{Column Identifiers: 144,700 Random Points}} \label{table07} \\

\hline \multicolumn{1}{|c|}{\textbf{ Column Number }} & \multicolumn{1}{c|}{\textbf{ Variable Name }} & \multicolumn{1}{c|}{\textbf{ Description }} \\ \hline 
\endfirsthead

\multicolumn{3}{c}%
{{\bfseries \tablename\ \thetable{} -- continued from previous page}} \\
\hline \multicolumn{1}{|c|}{\textbf{  Column Number  }} &
\multicolumn{1}{c|}{\textbf{  Name  }} &
\multicolumn{1}{c|}{\textbf{   Description   }} \\ \hline 
\endhead

\hline \multicolumn{3}{|r|}{{Continued on next page}} \\ \hline
\endfoot

\hline \hline
\endlastfoot

\rownumber & id2 & running index \\
\rownumber & x & x-Coordinate \\
\rownumber & y & y-Coordinate \\
\rownumber & z & z-Coordinate  \\
\rownumber & ra  & right ascension \\
\rownumber & dec & declination  \\
\rownumber & redshift & observed redshift \\
 \rownumber &  dCM/R\underscore Voronoi  & centroid  $\rightarrow$ point  (normalized) \\
 \rownumber & R\underscore Voronoi/dUniform\tablenotemark{a} & cell volume / total volume  \\ 
 \rownumber &  R\underscore Max  & Distance from sample to farthest vertex \\ 
\rownumber &  R\underscore Min  &  Distance from sample to nearest vertex  \\ 
\rownumber &  R\underscore Max  / R\underscore Min   &  Elongation \\ 
 \rownumber & cnWinners  & Class ID with most votes  \\
 \rownumber & volume  & Cell volume  \\\hline
\rownumber &  bb\underscore vol\underscore lev & Level ID BB(vol)\tablenotemark{b} \\ 
 \rownumber & bb\underscore vol\underscore blk & Block ID BB(vol)\tablenotemark{b}    \\
\rownumber &  bb\underscore den\underscore lev & Level ID BB(den)\tablenotemark{c}  \\
\rownumber & bb\underscore den\underscore blk  & Block ID BB(den)\tablenotemark{c} \\\hline 
\rownumber &  f55  & 0; but 1 if cell collision test fails \\
\rownumber &  fbad & 0; but 1 if boundary test fails  \\
\rownumber &  density\underscore akm  & KDE density  \\ 
\rownumber &  bandwidth\underscore akm  & KDE bandwidth    \\ 
\rownumber &  levels\underscore akm & KDE density level \\
\rownumber &  ID(Vor,+) )& Max-structure ID; HOP $\bf{f} =$ 1/volumel \\
\rownumber &  ID(Vor,-) & Min-structure ID; HOP $\bf{f} =$ 1/volume \\ 
\rownumber &  ID(AKM,+) & Max-structure ID; HOP $\bf{f} =$ 1/density\underscore akm \\ 
\rownumber &  ID(AKM,-) & Min-structure ID; HOP $\bf{f} =$ 1/density\underscore akm \\ 
\rownumber &  ID(SOM,+) & Max-structure ID; HOP $\bf{f} =$ cnWinners \\ 
\rownumber &  ID(SOM,-) & Min-structure ID; HOP $\bf{f} =$ cnWinners \\ 
\rownumber &  ID(BB(volume),+) & Max-structure ID; HOP $\bf{f} =$ n(blk)/volume; BB(vol)\tablenotemark{b} \\ 
\rownumber &  ID(BB(volume),-) & Min-structure ID; HOP $\bf{f} =$ n(blk)/volume; BB(vol)\tablenotemark{b} \\ 
\rownumber &  ID(BB(density),+) & Max-structure ID; HOP $\bf{f} =$ n(blk)/volume; BB(den)\tablenotemark{c}   \\ 
\rownumber &  ID(BB(density),-) & Min-structure ID; HOP $\bf{f} =$ n(blk)/volume; BB(den)\tablenotemark{c}  \\ 
\end{longtable}
    \begin{tablenotes}
    \item{(a) }{dUniform, the average distance between objects in the sample, equals 3.2e-3 redshift units.}
    \item{(b) }{Bayesian Block analysis based on Voronoi cell volume. }
    \item{(c) }{Bayesian Block analysis based on Voronoi cell density. }
    \end{tablenotes}
\end{center}


\subsection{Void Catalog Information: Electronic-Only Files}
\label{sec:void_catalog}

For each of the 3 data sets we have also constructed a flat ASCII file containing
identities and descriptions of the HOP voids. The names of these electronically
accessible files for the SDSS, MS and Poisson data are
\verb+delaunay_voids_sdss.txt+, 
\verb+delaunay_voids_ms.txt+ and 
\verb+delaunay_voids_poiss.txt+,
respectively.  All three files have the format given in Table \ref{table08}.
The identification number is a running index for the 458,173, 503,832,  and 
465,357 voids in the three cases, respectively.  The next two columns contain the
number of tetrahedra and galaxies in the void, followed by the void effective
radius in Eq. (\ref{eqn1}), convexity (see the caption to Fig. \ref{figure07}),
and distance from the augmented hull of the full data set.  The columns beginning
with number seven give the identities of the galaxies circumscribing
the void; these are the running index values $id2$ in Tables \ref{table05},
\ref{table06}, and \ref{table07}, respectively.  These rows contain a variable
number of galaxy IDs. All rows with less than the maximum number (25, 29, and 23
in the three cases, respectively) of galaxy IDs are are padded with zeros to
yield fixed record-length files. These lengths are 31, 35 and 29
in the three cases, respectively.

\begin{table}[htdp]
\caption{Delaunay Voids}\label{table09}
\begin{center}
\begin{tabular}{|  c |  l  | l |  }
\hline 
Column Number & Variable Name & Description \\
\hline
1 & VID & Void identification number \\
2 & $N_{\mbox{tet}}$ & Number of tetrahedra \\
3 & $N_{\mbox{gal}}$  & Number of galaxies \\
4 & $R_{\mbox{eff}}$  & Effective radius (Mpc)  \\
5 & $cvx$  & Convexity \\
6 & \verb+dist_min+ & Distance from hull \\
\ \ \ 7 ... & gid & Galaxy IDs \\
\hline
\end{tabular}
\end{center}
\label{table08}
\end{table}%

\section{Appendix B:  Zwicky Morphological Analysis of Topological Noise Effects}
\label{appendix_b}

The effect of sampling or measurement imperfections on data used to estimate a
distribution, especially in higher dimensions, is always much more complicated
than, say, for  the case of simple parameter estimation.  In the current spatial
statistics context, the various processes discussed in Section \ref{noise}
(and here, in the abuse of terminology described in that section,
termed \emph{noise}) can have several effects on the estimated structures.
Table \ref{table10} is a \emph{morphological box}, a device pioneered by
\citet{zwicky1948,zwicky} to facilitate complete investigation of parameter
spaces.  The first column (a) of this chart lists all six possible effects noise
of any kind may have on a specific structure, including creation of a new
structure and modification or destruction of an existing one. Column (b)
gives the accompanying net change in the number of structures.  Given the
``before'' size of a structure shown in the headings columns (c) - (f) 
the ``after'' size (with noise) is entered in the rows below.  The whole point
of this construct is to bring to light effects that might not be obvious at first
thought.  For example,  noise may actually bring about the apparent merger of
two structures into one (row E), e.g. turning  two small structures into one
medium structure as in box E(d).  A conclusion derivable from this matrix is 
that the common procedure of eliminating the smallest structures in the size
distribution may be only partially effective at de-noising.

\begin{table}[htdp]
\caption{{\bf Zwicky Morphological Matrix:} Effect of Noise on Numbers and Sizes of Structures}
\begin{center}
\begin{tabular}{ c | c | r | c |c|c|c|c|} 
\hline
&  (a)  & (b) & (c) & (d)  &  (e)  & (f)  \\ 
&{\bf Noise Effect } & $\delta$N &  {\bf Nonexistent} & {\bf Small}  & {\bf Medium} & {\bf Large} \\ \hline 
A & Create & +1 &  Small  &  ---   & ---   & ---   \\  \hline
B & Separate into Two & +1  & --- &   Small-  & Small  & Medium  \\  \hline
C & Reduce content & 0 & ---  &   Small-    &  Small  &Medium  \\  \hline
D & Increase content & 0 & --- &   Medium    & Large & Large+  \\  \hline
E & Merge &  -1  & ---  & Medium  & Large & Large+  \\ \hline
F & Destroy &  -1  & ---  &  Null  & Null  & Null  \\ \hline
\end{tabular}
\end{center}
\label{table10}
\tablenotetext{}{Note: The size indicators here are
only rough and nominal. The plus and minus signs are to be
interpreted as ``even larger" or ``even smaller," respectively.}
\end{table}%
\clearpage



\begin{thebibliography}{}

\bibitem[Abazajian et al.(2009)]{SDSS7}
Abazajian, K.N. et al. 2009, \apjs, 182, 543

\bibitem[Ahn et al.(2013)]{SDSS10}
Ahn, C., et al. 2012, arXiv:1307.7735 (Submitted to \apjs)

\bibitem[Aikio \& Mahonen(1998)]{aikio}
Aikio, J., \& Mahonen, P., 1998,
\apj, 497, 534.

\bibitem[Amendola et al.(1999)]{amendola1999}
Amendola, L., Frieman, J., 
\& Waga, I. 1999,
\mnras, 309, 465

\bibitem[Angulo et al.(2013)]{angulo}
Angulo, R., Chen, R., Hilbert, S., \& Abel, T.
2013, 
submitted to \mnras, arXiv:1309.1161v2


\bibitem[Aragon-Calvo et al.(2007)]{aragon_calvo_1}
Aragon-Calvo, M.,
Jones, B.,
van de Weygaert, R.,
\&
van der Hulst, J.,
2007,
\aap, 474, 315.

\bibitem[Aragon-Calvo et al.(2010a)]{aragon_calvo_2}
Aragon-Calvo, M.,
Platen, E.,
van de Weygaert, R.
\& Szalay, A. 2010a,
\apj, 723, 364.


\bibitem[Aragon-Calvo et al.(2010b)]{aragon_calvo_3}
Aragon-Calvo, M. A.,
van de Weygaerti, R.,
Araya-Melo, P. A.,
Platen, E.,
\&
Szalay A. S. 2010b,
\mnras, 404, L89.

\bibitem[Aragon-Calvo \& Szalay(2013)]{aragon_calvo_4}
Aragon-Calvo, M. A., \& Szalay A. S. 2013, \mnras, 428, 3409.


\bibitem[Aubert, Pichon \& Colombi(2004)]{aubert}
Aubert, D., Pichon. C and Colombi, S. 2004,
\mnras, 352, 376


\bibitem[Beygu et al.(2013)]{beygu}
Beygu, B., Kreckel, K., van de Weygaert, R., 
van der Hulst, J. M., van Gorkom, J. H. 2013,
\aj, 145, 120

\bibitem[Blanton et al.(2005)]{Blanton05}
Blanton, M. R. et al. 2005, \aj,  129, 2562

\bibitem[Bolejko et al.(2012)]{bolejko}
Bolejko, Krzysztof, Clarkson, Chris, Maartens, Roy, Bacon, David, Meures, Nikolai, Beynon, E. 2012,
\prl, 110, 021302


\bibitem[Bos et al.(2012)]{bos}
Bos, E., van de Weygaert, R., Dolag, K., \& Pettorino, V.(2012)
\mnras, 426, 440

\bibitem[Carlsson(2013)]{carlsson}
Carlsson, G, 2013
\verb+math.stanford.edu/?gunnar/actanumericathree.pdf+

\bibitem[Cautun(2011)]{cautun}
Cautun, M. \& van de Weygaert, R. 2011, 
The DTFE public software: The Delaunay Tessellation Field Estimator code, 2011
\verb+http://asterisk.apod.com/viewtopic.php?f=35&t=23588+

\bibitem[Chen et al.(2013)]{chen}
Chen, Y-C., Genovese, C., \& Wasserman, L. 2013,
``Uncertainty Measures and Limiting Distributions for 
Filament Estimation,''
arXiv:1312.2098v1


\bibitem[Ceccarelli et al.(2013)]{ceccarelli}
Ceccarelli, L., Paz, D., Lares, M., Padilla, N., \& 
Garcia Lambas, D. 2013,
``Clues on void evolution I: Large scale galaxy distributions around voids,''
arXiv:1306.5798v1

\bibitem[Chazal et al.(2014)]{chazal}
Chazal, F., Fasy, B., Lecci, F.,
Rinaldo, A., Singh, A. \& Wasserman, L. 2014,
``On the Bootstrap for Persistence Diagrams and Landscapes,''
arXiv:1311.0376v2




\bibitem[Cheng(2005)]{Cheng2005}
Cheng, T.P. 2005 ``Relativity, gravitation, and cosmology:
a basic introduction'', ISBN 9780198529576, Pub. Oxford University Press

\bibitem[Chincarini \& Rood(1976)]{chincarini_rood76}
Chincarini, G. \& Rood, H.J. 1976, ApJ, 206, 30

\bibitem[Cisewski et al.(2005)]{cisewski}
Cisewski, J., Croft, R., Freeman, P., Genovese, R., Khandai, N.,
Ozbek, M., \& Wasserman, L. 2014,
``Nonparametric 3D map of the IGM using
the Lyman-alpha forest,''
arxiv:1401.1867v1

\bibitem[Coil(2012)]{coil}
Coil, A. 2012, ``Large Scale Structure of the Universe,''
Volume 8., Section 9,
in Planets, Stars and Stellar Systems
Oswalt, T. D., ed. Springer: New York.
\verb+http://ned.ipac.caltech.edu/level5/March12/Coil/Coil9.html+


\bibitem[Colberg et al.(2008)]{colberg}
Colberg, J., Pearce, F. et al.
2008
\mnras, 387, 933.

\bibitem[Colless et al.(2001)]{Colless2001}
Colless, M.M. et al. 2001, \mnras, 328, 1039

\bibitem[Choi et al.(2010)]{Choi2010}
Choi Y. et al. 2010 \apjs, 109, 181

\bibitem[Croton et al.(2005)]{Croton2005}
Croton D.J. et al., 2005, \mnras, 356, 1155

\bibitem[Daley \& Vere-Jones(2003)]{daley}
Daley, D. \& Vere-Jones, D., 2003,
An Introduction to the Theory of Point Processes:\,
Vol. I: Elementary Theory and Methdos,
2nd Edition, Springer-Verlag: New York

\bibitem[D'Abrusco et al.(2012)]{abrusco}
DÕAbrusco, R., Fabbiano, G., Djorgovski, G., 
Donalek, C.,  Laurino, O. and G. Longo, G. 2012,
\apj, 755, 92


\bibitem[D'Aloisio \& Furlanetto(2007)]{aloisio}
D'Aloisio, A. \& Furlanetto, S. (2007),
\mnras, 382, 860

\bibitem[de Berg et al(1997)]{deberg}
de Berg, M., 
Cheong, O.,
van Kreveld, M.,
Ovvermars, M. 1997,
Computational Geometry: Algorithms and Applications,
Springer-Verlag: New York

\bibitem[Edelsbrunner (1987)]{edelsbrunner_1}
Edelsbrunner, H. 1987,
Algorithms in Combinatorial Geometry,
EATCS Monographs on Theoretical Computer Science,
Vo. 10, Springer-Verlag, New York

\bibitem[Edelsbrunner et al.(2002)]{edelsbrunner_2}
Edelsbrunner, H.,
Letscher, D., \& Zomorodian, A., 2002
Discrete Computational Geometry, 28, 511

\bibitem[Edelsbrunner \& Harer(2010)]{edelsbrunner_3}
Edelsbrunner, H. \& Harer, J., 2010,
Computational Topology,
American Mathematical Society: Providence, RI.

\bibitem[Einasto et al.(1989)]{einasto89}
Einasto, J., Einasto, M., \& Gramann, M.(1989)
\mnras, 238, 155


\bibitem[Einasto et al.(2011)]{einasto_gw}
Einasto, M.,  Liivam\"{a}gi, L.,Tempel, E., Saar, E.,
Tago, E., Einasto, P.,
 I. Enkvist, I.,
Einasto, J.,
Martinez, V.,
Hein\"{a}m\"{a}ki, P.,
\& Nurmi, P.(2011),
\apj, 736, 51


\bibitem[Einasto et al.(2012)]{einasto_a}
Einasto, M., Liivam\"{a}gi, L. J.,
Tempel, E., Saar, E.,
Vennik, J., Nurmi, P., Gramann, M.,
Einasto, J., Tago, E.,
HeinŠmŠki, P.,
 Ahvensalmi, A., \& Mart'nez, V.(2012)
\aap, 542, 36


\bibitem[Einasto et al.(2011)]{einasto_b}
Einasto, J., Suhhonenko, I., HŸtsi, G., Saar, E., 
Einasto, M., Liivam\"{a}gi, L. J., M\"{u}ller, V., Starobinsky, A. A., Tago, E., Tempel, E. 2011,
\aap 534, 128



\bibitem[Einasto et al.(2014)]{einasto_c}
Einasto, M., Lietzen, H., Tempel, E., 
Gramann, M., 
Liivam\"{a}gi, 
\& Einasto, J., 2014,
\aap 562, A87

\bibitem[Einastro(2014)]{EinastoBook2014}
Einasto, J. 2014, ``Dark Matter and Cosmic Web Story",
Edited by Jaan Einasto. Published by World Scientific Publishing Co. Pte. Ltd.,
2013. ISBN 9789814551052


\bibitem[Eisenstein \& Hut(1998)]{hop_1}
Eisenstein, D. \& Hut, P. 1998,
\apj,   498, 137

\bibitem[Elyiv et al.(2013)]{elyiv}
Elyiv, A., 
Karachentsev, I.,
Karachentseva, V.,
Melnyk, O.,
\& 
Makarov, D.(2013)
Astrophysical Bulletin 68, 1-12,
arXiv:1302.2369v2 

\bibitem[El-Ad et al.(1996)]{el_ad_1}
El-Ad, H., Piran, T. \& da Costa, L.(1996) \apj, L13

\bibitem[El-Ad(1997)]{el_ad_2}
El-Ad, H., (1997) ``The ($>$ Half) Empty Universe,
in From Quantum Fluctuations to Cosmological Structures,
ASP Conference Series, 126, 313.

\bibitem[El-Ad \& Piran(1997)]{el_ad_2a}
El-Ad, H., \&  Piran, T. (1997) 
\apj,   491, 421

\bibitem[El-Ad \& Piran(2000)]{el_ad_2b}
El-Ad, H., \&  Piran, T. (2000) 
\mnras, 313, 553


\bibitem[El-Ad et al.(1997)]{el_ad_3}
El-Ad, H., Piran,  T. \& da Costa, L. 1997, \mnras, 287, 790

\bibitem[Falck, Neyrinck \& Szalay(2012)]{falck2012}
Falck, B., Neyrinck, M. \& Szalay, A. 2012, 
Talk summary to appear in the Proceedings of the 13th Marcel Grossmann Meeting (MG13), Stockholm, July 2012,
arxiv:1309.4787N

\bibitem[Fasy et al.(2014)]{fasy}
Fasy, B., Lecci, F.,
Rinaldo, A., Wasserman, L., \& Singh, A. 2014,
``Statistical Inference for Persistent Homology:
Confidence Sets for Persistence Diagrams,''
arXiv:13037117v2

\bibitem[Forman(2002)]{forman}
Forman, R. 2002, S\'eminaire Lotharingien de Combinatoire 48, B48c

\bibitem[Gaite (2009)]{gaite}
Gaite, J., 2009, Journal of Cosmology and Astroparticle Physics, 1, 11

\bibitem[Gazis et al.(2010)]{GLW2010}
Gazis, P. R., Levit, C. \&  Way, M. J. 2010, PASP, 122, 1518



\bibitem[Gerber et al.(2010)]{gerber}
Gerber, S., Bremer, P. , Pascucci, V. , \& Whitaker, R., 2010,
IEEE Transactions on  Visualization and Computer Graphics,  
16, 1271

\bibitem[Gladders \& Yee (2000)]{gladders}
Gladders, M. \& Yee, H. 2000,
\aj, 120, 2148

\bibitem[Gott et al.(2008)]{gott}
Gott, J., Hambrick, D. et al. 2008,
\apj, 675, 16


\bibitem[Gregory \& Thompson(1978)]{gregory-thompson78}
Gregory, S. \& Thompson, L. 1978, \apj, 222, 784

\bibitem[Gyulassy \& Natarajan(2005)]{gyulassy}
Gyulassy, A. \& Natarajan, V.(2005),
Topology-based simplification for feature extraction from 3D scalar fields,
Visualization, 2005. VIS 05. IEEE


\bibitem[Hahn et al.(2007a)]{hahn_1}
Hahn, O., Porciani, C., Carollo, C.,  Marcella, C., \& Dekel, A. 2007,
\mnras, 375, 489

\bibitem[Hahn et al.(2007b)]{hahn_2}
Hahn, O., Carollo, C., Marcella, C., Porciani, C., \& Dekel, A. 2007,
\mnras, 381, 41

\bibitem[Hamaus et al.(2013)]{hamaus}
Hamaus, N., Wandelt, B., Sutter, P., Lavaux, G. \& Warren, M. 
arxiv:1307.2571v1

\bibitem[Hamaus et al.(2014)]{hamaus2014}
Hamaus, N., Sutter, P., \& Wandelt, B., 2014,
arxiv:1403.5499


\bibitem[Hidding, Shandarin \& van de Weygaert (2014)]{hidding}
Hidding, J., Shandarin, S., \& van de Weygaert, R. 2014,
\mnras, 437, 3442



\bibitem[Higuchi et al.(2011)]{higuchi}
Higuchi, Y., Oguri, M., \& Hamana, T.,
2011, \mnras, 432, 1021,
arxiv:1211.5966v2

\bibitem[Hogg(1999)]{hogg1999}
Hogg, D.W. 1999, arXiv:astro-ph/9905116v4

\bibitem[Holmberg (1969)]{holmberg}
Holmberg, E. 1969,
Arkiv F\"{o}r Astronomi,
5, 305

\bibitem[Hoyle \& Vogeley(2002)]{hoyle_vogeley}
Hoyle, F. \& Vogeley, M. 2002,
\apj, 566, 641

\bibitem[Ivezic et al.(2014)]{ivezic}
Ivezi\^{c}, Z., Connolly, A., VabderPlas, J., \& Gray, A.,
Statistics, Data Mining, and Machine Learning in Astronomy,
2014, Princeton University Press.

\bibitem[Jackson et al.(2010)]{jackson_higher_dimension}
Jackson, B., Scargle J., Cusanza, C., Barnes, D.,
Kanygin, R., Sarmiento, R., Subramaniam, S.,
Tzu-Wang, C. 2010,
Optimal Partitions of Data In Higher Dimensions,
in Proceedings of the 2010 Conference on Intelligent Data Understanding, eds: Ashok N. Srivastava, 
Nitesh Chawla, Philip Yu and Paul Melby,
NASA Ames Research Center: Mountain View


\bibitem[James et al.(2009)]{james2009}
James, J.B. et al. 2009, \mnras, 394, 454

\bibitem[Jasche et al.(2010)]{jasche2010}
Jasche, J. et al. 2010, \mnras, 406, 60

\bibitem[Jennings et al.(2013)]{jennings}
Jennings, E., Li, Y., \& Hu, W.(2013)
``The abundance of voids and the excursion 
set formalism,'' arXiv1304.6087

\bibitem[Joeveer et al.(1977)]{joeveer-einasto-tago77}
Joeveer, M., Einasto, J. \& Tago, M. 1977,
Tartu Astrof{\"u}{\"u}s.~Obs.~Preprint, Nr.~A-1, 45 

\bibitem[Joeveer et al.(1978)]{joeveer-einasto-tago78}
Joeveer, M., Einasto, J. \& Tago, M. 1978, \mnras, 185, 357

\bibitem[Joeveer \& Einasto(1978)]{joeveer-einasto78}
Joeveer, M. \& Einasto, J. 1978, in The large scale structure of the universe;
Proceedings of the Symposium, Tallin, Estonian SSR, September 12-16, 1977.
(A79-13511 03-90) Dordrecht, D. Reidel Publishing Co., 1978, p. 241-250

\bibitem[Kauffmann \& Fairall(1991)]{kauffmann_fairall}
Kauffmann, G. \& Fairall, A. 1991,
\mnras, 1991, 248, 313

\bibitem[Knebe et al.(2011)]{knebe2011}
Knebe, A., Knollmann,  et al. 2011,
\mnras, 415, 2293

\bibitem[Knebe et al.(2013)]{knebe}
Knebe, A., Pearce, F. et al. 2013,
\mnras, 435, 1618


\bibitem[Koenderink(1990)]{koenderink}
Koenderink, J. 1990,
Soliud Shape, MIT Press, Cambridge, Mass.

\bibitem[Kopylov \& Kopylova(2002)]{kopy}
Kopylov, A. \& Kopylova, F. 2002,
Astron. Astrophys. 382, 389.
Data at: 
\verb$http://vizier.cfa.harvard.edu/viz-bin/VizieR?-source=J/A+A/382/389$


\bibitem[Krause et al.(2013)]{krause}
Krause, E., Chang, T.-C., DorŽ, O., \& Umetsu, K. 2013,
\apj,  762, 20

\bibitem[Lavaux \& Wandelt(2010)]{lavaux2010}
Lavaux . G. \& Wandelt, B. (2010)
\mnras, 403, 1392

\bibitem[Lavaux \& Wandelt(2012)]{lavaux2012}
Lavaux . G. \& Wandelt, B. (2012)
\apj, 754, 109

\bibitem[Leistedt et al.(2013)]{leistedt2013}
Flaglets for studying the large-scale structure of the Universe,
Leistedt, B., Peiris, H., \& McEwen, J. 2013,
Proceedings of Wavelets and Sparsity XV, SPIE Optics and Photonics, 
arxiv:1308.5480

Lavaux, G. \& Wandelt, B. (2012)
\apj, 754, 109

\bibitem[Liivam\"{a}gi, Tempel \& Saar(2012)]{liivamagi}
Liivam\"{a}gi, L., Tempel, E., \& Saar, E. 2012,
\aap 539, A80

\bibitem[Longair(1978)]{longair78}
Longair, M.S. 1978, ``Personal View - The Large Scale Structure of the Universe,"
in The large scale structure of the universe; Eds. M. Longair and J. Einasto,
Proceedings of the Symposium, Tallin, Estonian SSR, September 12-16, 1977.
(A79-13511 03-90) Dordrecht, D. Reidel Publishing Co., 1978,

\bibitem[Longair \& Einastro(1978)]{Longair-Einasto78}
Longair, M.S. \& Einasto, J. 1978, ``The Large scale structure of the universe":
Symposium no. 79 held in Tallinn, Estonia, U.S.S.R., September 12-16, 1977,
Dordrecht, Holland ; Boston : D. Reidel Pub. Co., 1978.


\bibitem[Lowen \& Teich(2005)]{lowen}
Lowen, S. \& Teich, M. 2005,
Fractal-Based Point Processes,
John Wiley \& Sons: Hoboken, New Jersey


\bibitem[Martinez \& Saar(2002)]{martinez_saar}
Martinez, V. \& Saar, E. 2002,
Statistics of the Galaxy Distribution,
Chapman \& Hall/CRC, Boca Raton


\bibitem[Marzban \& Yurtsever(2011)]{marzban}
Marzban, C., Yurtsever, U. 2011,
Baby morse theory in data analysis,
in Proceedings of the 2011 workshop on Knowledge discovery, modeling and simulation,
ACM New York, NY

\bibitem[Milnor(1969)]{milnor}
Milnor, J., 1969,
Morse Theory, Annals of Mathematics Studies,
Princeton University Press.

\bibitem[M\"ueller et al.(2012)]{mueller2012}
M\"ueller, V., Hoffman, K. \&  Nuza, S.E. 2011, Baltic Astronomy, 20, 259

\bibitem[McBride et al.(2011)]{mcbride2011}
McBride, C.K. et al. 2011, \apj, 739, 85

\bibitem[Melchior et al.(2013)]{melchior2013}
Melchior, P., Sutter, P., Sheldon, E.,
Krause, E. \& Wandelt, B.
arxiv:1309.2045v1, submitted to \mnras

\bibitem[Neyrinck et al.(2005)]{neyrinck2005}
Neyrinck, M.C., Gnedin, N. \&  Hamilton, J. 2005, \mnras, 356, 1222


\bibitem[Neyrinck(2008)]{neyrinck2008}
Neyrinck, M.C., 2008, \mnras, 386, 2101

\bibitem[Neyrinck et al.(2009)]{neyrinck2009}
Neyrinck, M.C., Szapudi, I. \&  Szalay, A.S. 2009, \apj, 698, 90

\bibitem[Neyrinck et al.(2013)]{neyrinck2013}
Neyrinck, M.C., Aragon-Calvo, M., Jeong, D., \& Wang, X. 2013, 
arxiv:1309.6641

\bibitem[Nadathur \& Hotchkiss(2014)]{nadathur2014}
Nadathur, S. \& Hotchkiss, S. 2014,
\mnras, 440, 1248


\bibitem[Okabe et al.(2000)]{spatial_tessellations}
Okabe, A., Boots, B., Sugihara, K., \& Chiu, S. N. 2000, 
Spatial Tessellations: Concepts and Applications of Voronoi Diagrams,
John Wiley and Sons, Ltd., New York, Second Edition

\bibitem[Mu\~noz-Cuartas \& M\"uller(2012)]{FoF2012}
Mu\~noz-Cuartas, J.C. \& M\"uller, V. 2012, \mnras, 423, 1583

\bibitem[Pan et al.(2012)]{pan}
Pan, D., Vogeley, M., Hoyle, F., Choi, Y.-Y., Park, C. 
2012, 
\mnras, 421, 926

\bibitem[Pandey et al.(2013)]{pandey}
Pandey, B., White, S., Springel, V. \& Angulo, R. 2013,
\mnras, 435, 2968, arxiv:1301.3789v2


\bibitem[Paranjape et al.(2009)]{paranjape}
Paranjape, A., Lam, T., \& Sheth, R. 2011,
\mnras, 420, 1648

\bibitem[Park et al.(2012)]{park_great_wall}
Park, C., Choi, Y., Kim, J., Gott, J., Kim, S., Kim, K. 2012,
\apj,   759, 7


\bibitem[Park et al.(2013)]{park}
Park, C., Pranav, P., Chingangbam, P., van de Weygaert, R.,
Jones, B., Vegter, G., Kim, I., Hidding, J., \& Hellwing, W. 2013,
JKAS, 46, 125
arxiv:1307.2384

\bibitem[Patiri et al.(2013)]{patiri}
Patiri, S., Prada, F.,  Holtzman, J., Klypin, A. \& 
Betancort-Rijo, J. 2006, \mnras, 372, 1710

\bibitem[Peacock(1999)]{Peacock1999}
Peacock, J.A. 1999, ``Cosmologcal physics",
ISBN 9780521422703, Pub. Cambridge University Press


\bibitem[Peebles(1974)]{peebles_hierarchy}
Peebles, J. 2001, 
\apss, 31,403

\bibitem[Peebles(1980)]{peebles_2}
Peebles, J. 1980,
The large-scale structure of the universe,
Princeton University Press: Princeton

\bibitem[Peebles(1984)]{peebles_3}
Peebles, J. 1984,
Hierarchical Clustering, 
in Clusters and Groups of Galaxies,
Astrophysics and Space Science Library Volume, 111, 405


\bibitem[Peebles(2001)]{peebles_void}
Peebles, J. 2001, 
\apj, 557, 495


\bibitem[Platen et al.(2007)]{platen}
Platen, E., van de Weygaert, R., \& Jones, B. 2007,
\mnras, 380, 551

\bibitem[Platen et al.(2011)]{platen2011}
Platen, E., van de Weygaert, R., Jones, B.  Vegter, G., \& Aragon Calvo, M. 2011,
\mnras, 416, 2494

\bibitem[Preparata \& Shamos (1985)]{preparata}
Perparata, F. \& Shamos, M. 1985,
Computational Geometry: an Introduction,
Springer-Verlag, New York

\bibitem[Ricciardelli, Quilis and Varela(2014)]{ricciardelli}
Ricciardelli, E., Quilis, J. and Varela, J 2014,
\mnras, 440, 601,
arxiv:1402.2976

\bibitem[Rojas et al.(2004)]{rojas}
Rojas, R., Vogeley, M., Hoyle, F. \& Brinkman, J 2004,
\apj, 617, 50


\bibitem[Rykoff et al.(2014)]{redmap_3}
Rozo, E., Rykoff, E., Bartlett, J., \& Melin, J.
2014,
redMaPPer III: A detailed comparison of the Planck 2013
and SDSS DR8 RedMaPPer Cluster Catalogs,
arxiv:1401.7716

\bibitem[Rozo \& Rykoff(2013)]{redmap_2}
Rozo, E., Rykoff, E., 2013,
redMaPPer II: X-ray and SZ Performance Benchmarks 
for the SDSS Catalog,
arxiv:1303.3373

\bibitem[Rykoff et al.(2013)]{redmap_1}
Rykoff, E., Rozo, E., Busha, M., Cunha, C., 
Finoguenov, A., Evrard, A., Hao, J.,
Koester, B. P.; Leauthaud, A.; Nord, B..
Pierre, M., Sadibekova, T., Sheldon, E.,
\& Wechsler, R., 2013,
redMaPPer I: Algorithm and SDSS DR8 Catalog, 
arxiv:1303.3562


\bibitem[Scargle et al.(2013)]{scargle_vi}
Scargle, J., Norris, J., Jackson, B., and  Chiang, J. 2013,
\apj, 764, 167

\bibitem[Schaap(2007)]{schaap}
Schaap, W. 2007, 
DFTE: the Delaunay Tessellation Field Estimator,
Thesis:
\verb+http://dissertations.ub.rug.nl/faculties/science/2007/w.e.schaap/+


\bibitem[Schmidt et al.(2001)]{schmidt}
Schmidt, J., Ryden, B., \& Melott, A.(2001)
\apj, 546, 609

\bibitem[Schwarzschild(1982)]{schwarzschild82}
Schwarzschild, B. M., 1982,
Physics Today, 35, 17

\bibitem[Shandarin et al.(2004)]{shandarin2004}
Shandarin, S.F., Sheth, J.V. \&  Sahni, V. 2004, \mnras, 353, 162

\bibitem[Shectman et al.(1996)]{Shectman1996}
Shectman, S.A. et al. 1996, \apj, 470, 172

\bibitem[Skory et al.(2010)]{hop_2}
Skory, S., Turk, M., Norman, M., \& Coil, L. 2010, \apjs, 191, 43


\bibitem[Skrutskie et al.(2006)]{Skrutskie06}
Skrutskie, M.F. et al. 2005, \aj, 131, 1163


\bibitem[Snyder \& Miller(1991)]{snyder}
Snyder, D. \& Miller, M. 1991,
Random Point Processes in Time and Space,
Springer-Verlag: New York.

\bibitem[Sousbie(2013)]{sousbie_0}
Sousbie, T. 2013,
DisPerSE: robust structure identification in 2D and 3D, arXiv1302.6221S, 
cf. \verb+http://asterisk.apod.com/viewtopic.php?f=35&t=30806+
with software at
\verb+http://www2.iap.fr/users/sousbie/web/html/indexd41d.html?+

\bibitem[Sousbie(2011)]{sousbie_1}
Sousbie, T. 2011, \mnras, 414, 350

\bibitem[Sousbie et al.(2011)]{sousbie_2}
Sousbie, T., Pichon, C. \& Kawahara, H. 2011, \mnras, 414, 384


\bibitem[Springel(1999)]{Springel1999} 
Springel, V. 1999, 
PhD. Thesis: On the Formation and Evolution of Galaxies,
\verb+http://www.mpa-garching.mpg.de/~volker/+

\bibitem[Springel et al.(2005)]{Springel2005} Springel, V., et al. 2005, Nature,
435, 629

\bibitem[Strauss et al.(2002)]{Strauss02}
Strauss, M. A., et al. 2002 \aj, 124, 1810

\bibitem[Sutter et al.(2012)]{sutter}
Sutter, P., Lavaux, G., Wandelt, B. \& Weinberg, D.(2012) 
\apj,   761, 44


\bibitem[Tavasoli et al.(2013)]{tavasoli}
Tavasoli, S., Vasei, K. \& Mohayaee, R. 2013,
\aap 553, A15.

\bibitem[Tegmark et al.(2006)]{tegmark2006}
Tegmark, M., Eisenstein, D., et al.,  2006
Physical Review D, 74, 123507

\bibitem[Tempel et al.(2014)]{tempel}
Tempel, E., Tamm, A., Gramann, M., Tuvikene, T.,
 Liivam\"{a}ggi, L.,  Suhhonenko, I., Kipper, R., Einasto, M., \& Saar, E.
 2014,
arxiv:1402.1350v1

\bibitem[Thompson \& Gregory(2011)]{thompson-gregory2011}
Thompson, L.A. \& Gregory, S.A. 2011 arXiv:1109.1268

\bibitem[Tifft \& Gregory(1978)]{tifft-gregory1978}
Tifft, W.G. \& Gregory, S.A. 1978 ``Observations of the Large Scale Distribution
of Galaxies" in The large scale structure of the universe; Eds. M. Longair and J. Einasto,
Proceedings of the Symposium, Tallin, Estonian SSR, September 12-16, 1977.
(A79-13511 03-90) Dordrecht, D. Reidel Publishing Co., 1978, p. 267

\bibitem[Tully \& Fisher(1978)]{tully-fisher78}
Tully, R. \& Fisher, J. 1978 ``Nearby Small Groups of Galaxies,"
in The large scale structure of the universe; Eds. M. Longair and J. Einasto,
Proceedings of the Symposium, Tallin, Estonian SSR, September 12-16, 1977.
(A79-13511 03-90) Dordrecht, D. Reidel Publishing Co., 1978, p.31


\bibitem[Turk et al.(2011)]{turk}
Turk, M, Smith, B. D.,
Oshi, J., Kkory, S., Skillman, S., Abel, T., \& Norman, M.
\apjs, 192, 9

\bibitem[Valageas \&  Clerc(2012)]{valageas2012}
Valageas, P. \&  Clerc, N. 2012, arXiv:1205.4847

\bibitem[van de Weygaert et al.(2009)]{weygaert2009}
van de Weygaert, R., Jones, B., 
Platen, E., \& Aragon-Calvo, M. 2009,
in Sixth International Symposium on Voronoi Diagrams, 
2009, 3, arXiv:0912.3448

\bibitem[van de Weygaert et al.(2010)]{weygaert2010}
van de Weygaert, R., Vegter, G., 
Platen, E., Eldering, B., \& Kruithof, N. 2010, 
in SeventhInternational Symposium on Voronoi Diagrams, 
2010, 3, arXiv:0912.3448
arXiv:1006.2765

\bibitem[van de Weygaert et al.(2011a)]{weyg_a}
van de Weygaert, R., Kreckel, K., Platen, E., Beygu, B., van Gorkom, J. H., van der Hulst, J. M., Arag—n-Calvo, M. A., Peebles, P. J. E., Jarrett, T., Rhee, G., Kova\v{c}, K., Yip, C.-W. 2011,
The Void Galaxy Survey,
in Environment and the Formation of Galaxies: 30 years later, Astrophysics and Space Science Proceedings, ISBN 978-3-642-20284-1. 
Springer-Verlag Berlin Heidelberg, 2011, p. 17

\bibitem[van de Weygaert et al.(2011b)]{weyg_b}
van de Weygaert, R., Pranav, P., Jones, B., Bos, P., Vegter, G.,
Edelsbrunner, H., Teilaud, M., Hellwing, W., Park, C., Hidding, J., and Wintraecken, M. 2011,
Probing dark energy with alpha shapes and betti numbers,
arxiv:1110.5528


\bibitem[van de Weygaert et al.(2011c)]{weyg_c}
van de Weygaert, R.; Vegter, G.; Edelsbrunner, H.; 
Jones, B.; Pranav, P.; Park, C.; Hellwing, W.; Eldering, B.; 
Kruithof, N.; 
Bos, P.; Hidding, J.; Feldbrugge, J.; ten Have, E.; 
van Engelen, M.; Caroli, M.; \& Teillaud, M. 2011
Trans. Comput. Sci. XIV, pg. 60-101, special issue on Voronoi Diagrams and Delaunay Triangulation, eds. M. Gavrilova, C. Tan and M. Mostafavi (Springer)

\bibitem[Varela et al.(2012)]{varela}
Varela, J., Betancort-Rijo, J., Trujillo, I., \& Ricciardelli, E. 2012,
\apj, 744, 82



\bibitem[Way et al.(2011)]{WGS2011}
Way, M.J., Gazis, P.R. \&  Scargle, J.S. 2011, \apj, 727, 48

\bibitem[White(1979)]{white79}
White, S. D. M. 1979,
\mnras, 186, 145


\bibitem[York et al.(2000)]{York2000}
York, D.G., et al. 2000, \aj, 120, 1579

\bibitem[Yu \& Peebles(1969)]{yu_peebles}
Yu, J. \& Peebles, J. 1969,
\apj, 158,103


\bibitem[Zaninetti(2012)]{zaninetti}
Zaninetti, L. 2012,
Revista Mexicana de Astronom'a y Astrof'sica, 48, 209

\bibitem[Zeldovich(1978)]{zeldovich78}
Zeldovich, Y.B. 1978, ``The Theory of the Large Scale Structure of the Universe"
in The large scale structure of the universe; Eds. M. Longair and J. Einasto,
Proceedings of the Symposium, Tallin, Estonian SSR, September 12-16, 1977.
(A79-13511 03-90) Dordrecht, D. Reidel Publishing Co., 1978, p.409


\bibitem[Zomorodian(2005)]{zomorodian}
Zomorodian, A., 2005,
Topology for Computing, volume 16 of Cambridge Monographs on Applied and Computational
Mathematics. Cambridge University Press, New York, NY, 2005.


\bibitem[Zwicky(1948)]{zwicky1948}
Zwidky, F, 1948,
The Observatory, 68, 121


\bibitem[Zwicky(1957)]{zwicky}
Zwidky, F, 1957,
Morphological Astronomy,
Springer-Verlag: Berlin

\end{thebibliography}
\end{document}